\documentclass[manuscript,nonacm]{acmart}

\usepackage{amsmath,amsfonts}
\usepackage[ruled,vlined,noend,linesnumbered]{algorithm2e}
\usepackage{graphicx}
\usepackage{textcomp}
\usepackage{xcolor}
\usepackage{bbm}
\usepackage{url}
\usepackage{tikz}
\usepackage{comment}

\usepackage{subfigure}
\usepackage{makecell}
\usepackage[ruled,vlined]{algorithm2e}
\def\BibTeX{{\rm B\kern-.05em{\sc i\kern-.025em b}\kern-.08em
    T\kern-.1667em\lower.7ex\hbox{E}\kern-.125emX}}
    
\newcommand{\ourmethod}[0]{\textsc{certa}}
\newcommand{\ditto}{{Ditto}}

\SetKwFunction{FMain}{Main}

\newtheorem{example}{Example}

\newenvironment{myitemize}%
{\begin{list}{$\bullet$}{%
        \setlength{\labelsep}{5pt}%
        \setlength{\parsep}{2pt}%
        \setlength{\topsep}{2pt}%
        \setlength{\leftmargin}{10pt}%
        \setlength{\labelwidth}{20pt}%
        \setlength{\listparindent}{0pt}%
        \setlength{\itemsep}{.1\itemsep}}}
{\end{list}}

\begin{document}

\title{Effective Explanations for Entity Resolution Models}

\author{Tommaso Teofili}
\affiliation{\institution{Roma Tre University}
}
\email{tommaso.teofili@uniroma3.it}

\author{Donatella Firmani}
\affiliation{\institution{Sapienza University}
}
\email{donatella.firmani@uniroma1.it}

\author{Nick Koudas}
\affiliation{\institution{University of Toronto}
}
\email{koudas@cs.toronto.edu}

\author{Vincenzo Martello}
\affiliation{\institution{Roma Tre University}
}
\email{v.martello@inf.uniroma3.it}

\author{Paolo Merialdo}
\affiliation{\institution{Roma Tre University}
}
\email{paolo.merialdo@uniroma3.it}

\author{Divesh Srivastava}
\affiliation{\institution{AT\&T Chief Data Office}
}
\email{divesh@research.att.com}

\renewcommand{\shortauthors}{Teofili, et al.}

\newcommand{\myparagraph}[1]{\smallskip \noindent \textbf{#1. }}

\newcommand{\red}[1]{\textcolor{red}{#1}}

\begin{abstract}
Entity resolution (ER) aims at matching records that refer to the same real-world entity.  Although widely studied for the last 50 years, ER still represents a challenging data management problem, and several recent works have started to investigate the opportunity of applying deep learning (DL) techniques to solve this problem.  In this paper, we study the fundamental problem of explainability of the DL solution for ER. Understanding the matching predictions of an ER solution is indeed crucial to assess the trustworthiness of the DL model and to discover its biases.  We treat the DL model as a black box classifier and – while previous approaches to provide explanations for DL predictions are agnostic to the classification task – we propose the \ourmethod{} approach that is aware of the semantics of the ER problem. Our approach produces both saliency explanations, which associate each attribute with a saliency score, and counterfactual explanations, which provide examples of values that can flip the prediction. \ourmethod{} builds on a probabilistic framework that aims at computing the explanations evaluating the outcomes produced by using perturbed copies of the input records. We experimentally evaluate \ourmethod{}’s explanations of state-of-the-art ER solutions based on DL models using publicly available datasets,  and  demonstrate  the  effectiveness  of \ourmethod{} over recently proposed methods for this problem.
\end{abstract}

\maketitle


\section{Introduction}


Recent developments in Machine Learning (ML) and Deep Learning (DL)~\cite{pouyanfar2018survey} have had a profound impact on several research communities, especially computer vision \cite{wang2016database} and natural language understanding \cite{devlin2018bert}. ML/DL has also had considerable impact on data management research, yielding alternate proposals for, among other topics, query optimization, selectivity estimation, approximate query processing, and entity resolution~\cite{brunner2020entity,ebraheem2018distributed,mudgal2018deep,DBLP:journals/pvldb/0001LSDT20}.
Although DL models have demonstrated unparalleled prediction accuracy for very specific tasks, they are often criticized as offering predictions without any intuition or rationale~\cite{lapuschkin2019unmasking}. 

Entity Resolution (ER) is the task that aims at matching records that refer to the same real-world entity. Although widely studied for the last 50 years~\cite{fellegi1969theory}, ER still represents a challenging data management problem. Recent works have investigated the application of DL techniques to solve the ER problem~\cite{brunner2020entity,ebraheem2018distributed,mudgal2018deep,DBLP:journals/pvldb/0001LSDT20}. A typical application of an ML model to the ER problem involves the training of a classifier, possibly a deep neural network, for this problem~\cite{brunner2020entity,ebraheem2018distributed,mudgal2018deep,DBLP:journals/pvldb/0001LSDT20}. Given a set of training data and associated labels (match or non-match), a classifier is trained to solve a binary classification problem. Subsequently given a pair of records, the records are suitably encoded and the classifier yields a binary prediction for the pair. As with any classification problem, it is assumed that future data follow the same distribution as that of the training data set. The ML classification models applied to this problem typically apply either traditional SVM~\cite{christen2008automatic},  LSTM architectures~\cite{ebraheem2018distributed,mudgal2018deep} or deep transformer architectures like BERT \cite{brunner2020entity,DBLP:journals/pvldb/0001LSDT20}. Several recent approaches have demonstrated impressive prediction accuracy for the ER problem~\cite{primpeli2020profiling,barlaug2021neural}.


Since DL models typically do not come with any explanations providing reasons for their predictions, an active research area has been the exploration of techniques to offer {\em explainable} predictions revealing the process the DL network followed to reach its decision~\cite{guidotti2018survey}. Explanations represent an effective way to debug the system and are fundamental to trust its decisions, as they aim to provide the rationale behind a classifier's predicted outcome.
For example, explanations are useful in situations where an ML classifier for ER makes wrong \emph{predictions} (either classifies a match as non-match or vice-versa), as well as they can assist to check whether a classifier is making correct predictions for sound reasons.

Figure~\ref{fig:example} reports some records from the \emph{Abt-Buy} dataset, a popular benchmark for ER~\cite{mudgal2018deep}. Figure~\ref{fig:example-outcome} shows the predictions obtained for three such record pairs by three ER systems based on DL, namely DeepER~\cite{ebraheem2018distributed}, DeepMatcher~\cite{mudgal2018deep}, and {\ditto}~\cite{DBLP:journals/pvldb/0001LSDT20}. The three pairs are in match, but all the three systems make mistakes on one of them (even {\ditto}, which performs very well, with $F1 \simeq 0.91$, on that dataset). Observe that the pairs in fact are rather similar: having explanations about the wrong predictions could help understand the roots of the misclassifications and improve the performance of the DL systems for ER. 
Popular approaches to provide an explanation for an ML classifier output are based on \emph{saliency} and \emph{counterfactual} explanation methods~\cite{arya2019one,martens2014explaining}.

\begin{figure*}
\scriptsize
\centering
\begin{minipage}[l]{\columnwidth}
   \subfigure[Abt\label{subtab:abt}]{
    \setlength{\tabcolsep}{0.5em}
    \begin{tabular}{|c|l|l|l|}
    \cline{2-4}
    \multicolumn{1}{c|}{} & \textbf{Name$_{Abt}$} & \textbf{Description$_{Abt}$} & \textbf{Price$_{Abt}$} \\     
    \cline{2-4} \hline
    $u_1$  & sony bravia theater black & sony bravia theater & NaN\\
     & micro system davis50b & black micro... & \\
     \hline
    $u_2$  & altec lansing inmotion  & altec lansing inmotion ipod  & NaN  \\
     & portable audio system ... & portable audio system im600usb... &   \\
     \hline
    $u_3$ & sony 19 ' bravia m-series & sony 19 ' bravia m-series silver & NaN \\
     & silver lcd flat panel hdtv ... & lcd flat panel hdtv ... &  \\
    \hline
    \end{tabular}
   }
\end{minipage}  
\begin{minipage}[r]{\columnwidth} 
  \subfigure[Buy\label{subtab:buy}]{
    \setlength{\tabcolsep}{0.5em}
    \begin{tabular}{|c|l|l|l|}
    \cline{2-4}
    \multicolumn{1}{c|}{} & \textbf{Name$_{Buy}$} & \textbf{Description$_{Buy}$} & \textbf{Price$_{Buy}$} \\     
    \cline{2-4} \hline
    $v_1$  & sony bravia dav-is50 / b  & dvd player , 5.1 speakers & NaN \\
    & home theater system  & 1 disc ( s ) progressive ... & \\
    \hline
    $v_2$ & altec lansing inmotion  &  & NaN \\
    & im600 portable audio ... &  &  \\
    \hline
    $v_3$ & sony bravia m series ... & 19 ' atsc , ntsc 16:9 1440 x 900 ... & 379.72 \\  
    \hline
    \end{tabular}
   }
\end{minipage}  

\caption{Sample records from the Abt-Buy dataset.} \label{fig:example} 
\end{figure*}

\begin{figure}
    \centering
    \scriptsize
    \begin{tabular}{|c|c|c|c|c|}
    \hline
    \textbf{Input} & \textbf{Ground-Truth} & \textbf{{\ditto}} & \textbf{DeepMatcher} & \textbf{DeepER} \\ \hline \hline
    $\langle u_1, v_1 \rangle$ & Match & Match & Match & \textbf{Non-Match}   \\
    &  & (0.98) & (0.71) & \textbf{(0.01)}   \\
    \hline
    $\langle u_2, v_2 \rangle$ & Match & Match & \textbf{Non-Match} & Match \\
    & & (0.93) & \textbf{(0.16)} & (0.69) \\
    \hline
    $\langle u_3, v_3 \rangle$ & Match & \textbf{Non-Match} & Match & Match  \\
    & & \textbf{(0.002)} & (0.73) & (0.89)  \\
    
    \hline
    \end{tabular}

    \caption{ER predictions performed by different DL systems on three pairs of the records from Figure~\ref{fig:example}. In brackets, the matching \emph{score} of the system: for all the systems, $score \in [0, 1]$, and $score > 0.5$ corresponds to Match.}
    \label{fig:example-outcome}
\end{figure}


\begin{figure*}
\footnotesize
\centering
    \begin{tabular}{|l|l|l|l|l|}
    \cline{2-5}
    \multicolumn{1}{c|}{} & \multicolumn{4}{c|}{\textbf{Explanation (Saliency)}} \\ \hline
    \textbf{ER System on tuple} & \textbf{\ourmethod} & \textbf{Mojito}  &  \textbf{LandMark} & \textbf{SHAP} \\ \hline \hline
    DeepER on $\langle u_1, v_1 \rangle$ & $\text{Description}_{Abt}$, $\text{Name}_{Buy}$ & $\text{Name}_{Buy}$, $\text{Name}_{Abt}$ & $\text{Name}_{Abt}$, $\text{Description}_{Buy}$ & $\text{Description}_{Abt}$, $\text{Price}_{Buy}$  \\
    DeepMatcher on $\langle u_2, v_2 \rangle$  & $\text{Description}_{Buy}$, $\text{Name}_{Buy}$ & $\text{Description}_{Abt}$ & $\text{Description}_{Abt}$, $\text{Price}_{Abt}$ & $\text{Description}_{Abt}$, $\text{Name}_{Abt}$ \\
    {\ditto} on $\langle u_3, v_3 \rangle$ & $\text{Description}_{Buy}$, $\text{Name}_{Buy}$ & $\text{Description}_{Buy}$, $\text{Price}_{Buy}$ & $\text{Description}_{Abt}$, $\text{Name}_{Abt}$ & $\text{Price}_{Abt}$, $\text{Name}_{Buy}$ \\
    \hline
    \end{tabular}
\caption{Saliency explanations generated with different techniques for the wrong predictions of Figure~\ref{fig:example-outcome}.
\label{fig:example-explanations}} 
\end{figure*}

\begin{figure}[t]
\footnotesize
\centering
    \setlength{\tabcolsep}{0.43em}
    \begin{tabular}{|l|c||c|c|c|c|}
    \cline{2-6}
    \multicolumn{1}{c|}{} & \multicolumn{5}{c|}{\textbf{Matching Score}} \\ \hline
    \textbf{ER  System  on  tuple} & \textbf{Original}& \textbf{\ourmethod{}} & \textbf{Mojito}  & \textbf{LandMark} & \textbf{SHAP} \\ \hline \hline
    DeepER on $\langle u_1, v_1 \rangle$ & 0.01 & 0.35 & 0.03 & 0.15 & 0.02 \\
    DeepMatcher  $\langle u_2, v_2 \rangle$ & 0.16 & 0.97 & 0.17 & 0.24 & 0.16 \\
    {\ditto} on $\langle u_3, v_3 \rangle$ & 0.002 & 0.99 & 0.15 & 0.008 & 0.002 \\
    \hline
    \end{tabular}
\caption{Inspecting the faithfulness of saliency explanations generated with different techniques.\label{fig:example-faithfulness}} 
\end{figure}

\myparagraph{Saliency methods} These methods explain the prediction of the classifier by assigning a \emph{saliency} score to each feature in the specific prediction input. This way, the features that influence the predicted outcome the most can be identified.
%
%

In the context of explaining the results of a classifier for ER, saliency methods aim at identifying the most influential attributes in an input pair, with respect to the predicted outcome. In the example of Figure~\ref{fig:example-outcome}, a saliency method should identify which attributes in the pair $\langle u_3, v_3 \rangle$ are influencing {\ditto} predict it as a non-match the most.
Notable examples of saliency methods are LIME~\cite{ribeiro2016should} and SHAP~\cite{DBLP:conf/nips/LundbergL17}, which were conceived for generic classification tasks on textual data and images, ignoring the semantics of the problem the classifier is used to solve.
Mojito~\cite{di2019interpreting} and LandMark~\cite{DBLP:conf/edbt/BaraldiBP021} represent adaptations of these methods specifically tailored for the ER task. Saliency explanation methods are sometimes also referred as feature attribution methods in literature.

\myparagraph{Counterfactual explanations} These methods help understanding the behavior of the system by providing modified copies of the original input that lead to a different predicted outcome than the original prediction.
In our example, a counterfactual explanation can help answering the question \emph{"how the pair $\langle u_3, v_3 \rangle$ should be (minimally) changed in order to make {\ditto} predict it as a match?"}.
Counterfactual explanations for ER systems, to the best of our knowledge, have not been explored at all in the literature, while there are several task agnostic  methods, including DiCE~\cite{DBLP:conf/fat/MothilalST20}, and the counterfactual versions of LIME and SHAP, \emph{LIME-C} and \emph{SHAP-C}~\cite{DBLP:journals/adac/RamonMPE20}. 

It has been observed that saliency and counterfactual explanation methods are different but complimentary methods to be used to best evaluate causality aspects of a classifier prediction~\cite{kommiya2021towards}. Saliency methods align well with the notion of \emph{necessity}, while counterfactual explanation methods align with the notion of \emph{sufficiency}~\cite{watson2021local}.

This paper presents {\ourmethod}, an original method that provides both saliency and counterfactual explanations for ER systems. {\ourmethod} considers specific characteristics of the ER task, and builds on the sound theoretical framework developed by Watson {\em et al.}~\cite{watson2021local}, which frames the concepts of probability of necessity and probability of sufficiency in the context of explanations. 


We focus on \emph{attribute-level} explanations because they align well with the way input data is structured and understood by users of structured relational databases. Attribute-level explanations are a natural choice for ER over such structured data sources where records are treated as a composition of attributes and primarily compared attribute-wise. Different ER systems have been designed for either explicitly capturing attribute-level information (e.g., \emph{DeepMatcher}~\cite{mudgal2018deep}) or injecting attribute-level domain knowledge (see Sect. 3.1 and 3.3 in the \ditto{} paper~\cite{DBLP:journals/pvldb/0001LSDT20}).


While previous proposals \cite{di2019interpreting, DBLP:conf/edbt/BaraldiBP021} represent interesting attempts to provide explanations to ER systems, they lack a theoretical foundation and the effectiveness of their explanations is limited.


Figure~\ref{fig:example-explanations} shows the saliency explanations generated by \ourmethod{}, Mojito, LandMark and SHAP for the wrong predictions of Figure~\ref{fig:example-outcome}. Observe that the four approaches produce different explanations. For example, \ourmethod{} indicates that the most influential attributes for the DeepER results are Description from the Abt table and Name from the Buy table (denoted as Description$_{Abt}$ and Name$_{Buy}$, respectively), while Mojito identifies Name$_{Buy}$ and Name$_{Abt}$. Similarly, Figure~\ref{fig:example-computerfactual} shows counterfactual explanations generated by \ourmethod{} and by DiCE for the prediction of DeepER on the pair $\langle u_1, v_1 \rangle$. For each method, we report in boldface the values of the generated explanation that should flip the prediction (from non-match to match). Note that the different explanations provide contrasting results.  

Given such a diversity of results, one may wonder which explanation is the most faithful to the actual behavior of the ER system. For saliency methods, one way to evaluate the effectiveness of an explanation consists of computing a new prediction using as input an altered pair, where the values of the attributes indicated by the saliency method are copied into the other tuple. For example, in evaluating the faithfulness of LandMark, copying the value of  Name$_{Abt}$ into Name$_{Buy}$, and the value of Description$_{Buy}$ into Description$_{Abt}$. As the tuples have been made more similar by the attributes that most influenced the decision, it is expected that the matching score of the classifier increases. Similarly, for a counterfactual explanation it is possible to check how the values suggested by the explanation method change the prediction.   

Figure~\ref{fig:example-faithfulness} shows the original matching on the original input pairs and those obtained by modifying the input pairs according to the explanations of Figure~\ref{fig:example-explanations}. For all the methods but \ourmethod{}, the matching scores do not change significantly, even if the tuples have been made more similar by following the insights of the explanations. Apparently, the saliency computed by these explanation methods does not reflect the importance of the attributes for the decisions of the ER systems. In contrast, the explanation generated by \ourmethod{} changes the matching score a lot. Similarly,  Figure~\ref{fig:example-computerfactual} reports the matching score of DeepER on the pair modified as suggested by the explanation. Also in this case, it is easy to observe that \ourmethod{} produces a more effective explanation, which actually forces the system to flip the prediction (since the resulting  matching score is greater than $0.5$). 

In Section~\ref{sec:exp} we provide results of an extensive evaluation that demonstrates the superiority of \ourmethod{} in a wider experimental setting.

\myparagraph{Contributions}
We make the following contributions in the context of providing explanations for ER models:
\begin{myitemize}
\item We present the \ourmethod{} algorithm, which can exploit the semantics of the ER problem to provide saliency and counterfactual explanations that are quantitatively effective with respect to previous approaches.
\item We introduce the first counterfactual explanation technique for ER classifiers.
\item We present a principled framework based on the notions of probability of necessity and sufficiency and lattice structures.
\item We experimentally evaluate \ourmethod{}'s explanations of state-of-the-art ER solutions based on DL models using publicly available datasets, and demonstrate the effectiveness of \ourmethod{} over recently proposed methods for this problem.\footnote{The source code of \ourmethod{} is available at \url{https://github.com/tteofili/certa}.}
\end{myitemize}

\myparagraph{Paper outline}
Section~\ref{sec:related} discusses related work. Section~\ref{sec:problem} introduces the problem statement. Section~\ref{sec:lattices} describes our approach to efficiently compute saliency and counterfactual explanations. Section~\ref{sec:exp} presents the experimental evaluation that we have conducted. Section~\ref{sec:conclusion} discusses concluding remarks and future work.

\begin{figure*}[ht]
\scriptsize
\centering
    \setlength{\tabcolsep}{0.3em}
    \begin{tabular}{|c|c|l|l|c|l|l|c|}
    \cline{3-8}
    \multicolumn{2}{c|}{} 
     & \multicolumn{6}{c|}{\textbf{Counterfactual explanation}}\\
    \cline{2-8}
    \multicolumn{1}{c|}{} 
     & \textbf{Matching Score} & $\textbf{Name}_{Abt}$ & $\textbf{Description}_{Abt}$ & $\textbf{Price}_{Abt}$ & $\textbf{Name}_{Buy}$ & $\textbf{Description}_{Buy}$ & $\textbf{Price}_{Buy}$\\ \hline \hline
    CERTA & 0.54 & \makecell{sony bravia theater black \\ micro system davis50b} & \makecell{\textbf{denon 5-disc cd auto changer} \\ \textbf{dcm290 cd-r/rw playback advanced ...}} & NaN &
    \makecell{sony bravia dav-is50 / b \\ home theater system } & \makecell{"dvd player , 5.1 speakers\\ \ 1 disc ( s ) progressive scan...} & NaN \\
    \hline
    DiCE & 0.34 & \makecell{\textbf{lg 14 ' washer and }\\ \textbf{ dryerred pedestal ...}} & \makecell{sony bravia theater black micro\\ system davis50b 5.1-channel surround } & NaN &
    \makecell{\textbf{canon pixma mx700}\\ \textbf{multifunction photo ...}} & \makecell{\textbf{lithium ion ( li-ion )}\\ \textbf{8.4 v dc photo battery}} & NaN \\
    \hline
    \end{tabular}
\caption{Counterfactual explanations by  \ourmethod{} and \emph{DiCE} for the DeepER prediction on $\langle u_1, v_1 \rangle$: the values of the attributes identified by each counterfactual explanation method are highlighted in \textbf{bold}. The matching score is computed modifying the original pair using the values suggested by the counterfactual explanation. The original matching score is equal to $0.01$ (Non-Match), as reported in Figure~\ref{fig:example-outcome}.} \label{fig:example-computerfactual} 
\end{figure*}


\section{Related works}
\label{sec:related}



Much recent research has been conducted in the context of \emph{explainable AI}~\cite{guidotti2018survey}. Explanation systems can be divided into different categories, in particular we focus on saliency and counterfactual explanation systems. Saliency explanation systems describe the relationship between input features and the output of a model, for example providing a relevance score for each feature. One of the best known systems is LIME~\cite{ribeiro2016should}, which aims at explaining the prediction of any classifier for text, images or tabular data. Another explanation system, called SHAP~\cite{DBLP:conf/nips/LundbergL17}, develops a saliency explanation scheme based on game theoretic concept of Shapley values. All such methods can be applied in principle to any classification task, including ER. However, in the case of ER the classification task takes as input pairs of records rather than a single record (e.g., as in image classification tasks) and using the mentioned general purpose explanation methods may not be desirable. We refer the reader to~\cite{wang2018explaining,thirumuruganathan2019explaining} for further discussion on the problem of providing explanation methods for the ER task and for data integration in general.

More recently, new explanation systems have been proposed for the ER task, namely, Mojito~\cite{di2019interpreting},  ExplainER~\cite{DBLP:conf/icde/EbaidTAEO19} and LandMark~\cite{DBLP:conf/edbt/BaraldiBP021}.
Mojito~\cite{di2019interpreting} provides an adaptation of a general purpose explanation method -- that is, LIME~\cite{ribeiro2016should} -- on ER models. Mojito introduces two specific operations: ``Mojito pre-processing'', which transforms a record pair to a string representation, and ``LIME COPY'', which generates new record pairs in conjunction with the standard ``DROP'' operator provided by LIME.
LandMark~\cite{DBLP:conf/edbt/BaraldiBP021} provides a further adaptation of LIME to the specific setting of Entity Resolution. It internally generates two explanations for each record pair, each one explaining the classifier (with LIME) when the other record is kept unchanged.
ExplainER~\cite{DBLP:conf/icde/EbaidTAEO19} provides a unified graphical user interface to identify representative pairs to understand the model's behavior and identify attributes that are overall more influential. In the back-end, ExplainER can plug-in different general purpose explanation systems (including, LIME~\cite{ribeiro2016should} and Anchors~\cite{ribeiro2018anchors}) by modeling the ER task as a binary text classification task.
We note that Mojito, ExplainER, LandMark consist of more or less advanced adaptations of general purpose methods to the ER task, and do not provide any new explanation method. 
A complementary approach to explainable ER was recently proposed by SystemER~\cite{DBLP:journals/pvldb/QianPS19}. Even though SystemER is not an explanation system, it enables the user to learn an inherently explainable ER model, with human-comprehensible rules and the desired level of quality, by involving expert humans in the loop. 

Several counterfactual explanation approaches have been developed \cite{DBLP:journals/access/StepinACP21, DBLP:journals/corr/abs-2010-10596}. For the sake of this work, we consider counterfactual explanation methods that can treat the ER classifier as a black box function. 
In this context, model agnostic counterfactual explanation approaches that can be adapted to the ER task include DiCE~\cite{DBLP:conf/fat/MothilalST20}, LIME-C and SHAP-C~\cite{DBLP:journals/adac/RamonMPE20}, which we adopt as baselines.
Other interesting counterfactual frameworks that need access to the inner workings of the classifier include \cite{DBLP:journals/corr/abs-1711-00399, DBLP:conf/pkdd/LooverenK21}.
To the best of our knowledge no counterfactual explanation methods specifically designed for the ER setting exist yet.

\section{Foundations and problem statement}
\label{sec:problem}

We refer to real-world objects (e.g., products, persons, organizations) as \emph{entities} and to structured entity descriptions as \emph{records}. Given two sets of records, $U$ and $V$, ER consists of identifying all the record pairs $u,v \in U \times V$ that refer to the same entity. We say that record pairs referring to the same entity are \emph{matching}, and denote as $E^+ \subseteq U \times V$ the set of matching record pairs in the ground truth. Analogously, we refer as $E^- = (U \times V) \setminus E^+$ to the set of non-matching record pairs. 
We assume that records $u \in U$ have attributes $A_U=\{a_{U_1}, a_{U_2} \dots, a_{U_h}\}$ and, similarly, records $v \in V$ have attributes $A_V=\{a_{V_1}, a_{V_2} \dots, a_{V_k}\}$, therefore $U$ and $V$ may have different schemas. We refer to the value of the $i$-th attribute of a record $r \in U$ (resp. $V$) as $r[a_{U_i}]$ (resp. $r[a_{V_i}]$), with $a_{U_i} \in A_U$ (resp. $a_{V_i} \in A_V$). 



\myparagraph{ER Explanations} We are interested in providing explanations for a model $M$ solving ER as a binary classification problem. We refer as $M(\langle u,v \rangle)$ to the function learned by the model $M$. Such a function ought to be $\mathbb{T}$ (true) if $(u,v) \in E^+$, and $\mathbb{F}$ (false) otherwise, but can make mistakes if the model is not perfect. The model $M$ can be trained with a subset $T^+ \cup T^-$ of the ground truth $E^+ \cup E^-$ (with $T^+ \subseteq E^+$ and $T^- \subseteq E^-$), or can be unsupervised.

A \emph{local} explanation aims at describing the behavior of $M$ for a single prediction $M(\langle u, v \rangle) = y$.
A post-hoc explanation method involves an auxiliary method to explain $M$ after it has been trained. 
We distinguish two types of post-hoc local explanations, \emph{saliency} explanations and \emph{counterfactual} explanations, as follows:

\begin{myitemize}
    \item 
    A \emph{saliency} explanation for ER assigns an importance score to each attribute 
    $a \in A_U \cup A_V$,
    for a prediction $M(\langle u, v \rangle) = y$. The saliency score aims at capturing the contribution of the attribute to the predicted value.
    \item
    A \emph{counterfactual} explanation provides input samples that change a prediction to a desired outcome.
    Same as for the saliency explanations, we focus on providing attribute based counterfactual explanations.
    A counterfactual explanation for $M(\langle u, v \rangle) = y$ 
    consists of a pair $\langle u^{\prime}, v^{\prime} \rangle$, that is equal to $\langle u, v \rangle$ except for one or more attribute values and results in $M(\langle u^{\prime}, v^{\prime} \rangle) = \overline{y}$.
\end{myitemize}

Similarly to other popular explanation techniques, in order to generate saliency and counterfactual explanations for a prediction $M(\langle u,v \rangle) = y$, we resort to the notion of \emph{perturbation}, which
consists of assessing how altering (perturbing) portions an input sample affects the corresponding prediction yielded by the model. In particular, we evaluate the influence that attributes of the input pair have on the prediction by verifying if perturbing their values yields a \emph{flip} in the prediction outcome. 

Our approach to generate the perturbations is based on the following intuitions.
%
Consider the prediction
$M(\langle u,v \rangle) = \mathbb{F}$, with $u \in U$ and $v \in V$, for which we want to generate an explanation. Let $w \in U$ be a record such that $M(\langle w, v \rangle) = \mathbb{T}$, that is, $w$ and $v$ are a match according to M. As depicted in Figure~\ref{fig:trans2}, if we  progressively copy attribute values from $w$ to $u$, deriving a $u'$, increasingly making $u'$ more similar to $w$ based on their content, at some point the prediction of the model will flip, declaring $u'$ and $v$ to be a match. Repeating the same procedure for many records $w \in U$ produces evidence of the influence that attributes and set of attributes have on the input prediction. A similar argument can be formulated for the case of two records $u$ and $v$ that are predicted as a match by the model, i.e., $M(\langle u,v \rangle) = \mathbb{T}$, as depicted in Figure~\ref{fig:trans1}. Analogously, we can derive the sets of attributes that if their corresponding attribute values are altered the pair becomes a non-match in a consistent manner.

The above intuitions are formalized by the concepts of \emph{open triangle} and  \emph{open triangle perturbations}, which are the building blocks for our probabilistic definition of saliency explanation and counterfactual explanation.  





\begin{figure}
\centering
\subfigure[$M(\langle u,v\rangle) = \mathbb{F}$, $M(\langle w,v \rangle) = \mathbb{T}$]{
\begin{tikzpicture}
  \node (u) at (-1.5,0) {$\scriptstyle u \in U$};
  \node (v) at (1.5,0) {$\scriptstyle v \in V$};
  \node (w) at (0,2) {$\scriptstyle w \in U$};
  \draw[preaction={draw=red, -,line width=6pt}] (u) -- (v);
  \draw[preaction={draw=green, -,line width=6pt}] (v) -- (w);
\end{tikzpicture}
}
\hspace{1 cm}
\subfigure[Making the perturbed version of $u$, denoted $u'$, more similar to $w$ by copying values from $w$ to $u$ triggers  $M(\langle u',v\rangle)=\mathbb{T}$.]{
\begin{tikzpicture}
  \node (u) at (-1.5,0) {$\scriptstyle u' $};
  \node (v) at (1.5,0) {$\scriptstyle v \in V$};
  \node (w) at (0,2) {$\scriptstyle w \in U$};
  \draw[preaction={draw=green, -,line width=6pt}] (u) -- (v);
  \draw[preaction={draw=green, -,line width=6pt}] (v) -- (w);
  \draw[black, dashed, ->, inner sep=0pt] (w) -- (u);
\end{tikzpicture}
}
\caption{Perturbation on a non-matching pair $\langle u, v \rangle$.}
\label{fig:trans2}
\end{figure}

\begin{figure}
\centering
\subfigure[$M(\langle u,v \rangle) = \mathbb{T}$, $M(\langle w,v\rangle) = \mathbb{F}$]{
\begin{tikzpicture}
  \node (u) at (-1.5,0) {$\scriptstyle u \in U$};
  \node (v) at (1.5,0) {$\scriptstyle v \in V$};
  \node (w) at (0,2) {$\scriptstyle w \in U$};
  \draw[preaction={draw=green, -,line width=6pt}] (u) -- (v);
  \draw[preaction={draw=red, -,line width=6pt}] (v) -- (w);
\end{tikzpicture}
}  
\hspace{1 cm}
\subfigure[Making the perturbed version of $u$, denoted $u'$, more similar to $w$ by copying values from $w$ to $u$ triggers $M(\langle u',v \rangle)=\mathbb{F}$.]{
\begin{tikzpicture}
  \node (u) at (-1.5,0) {$\scriptstyle u'$};
  \node (v) at (1.5,0) {$\scriptstyle v \in V$};
  \node (w) at (0,2) {$\scriptstyle w \in U$};
  \draw[preaction={draw=red, -,line width=6pt}] (u) -- (v);
  \draw[preaction={draw=red, -,line width=6pt}] (v) -- (w);
  \draw[black, dashed, ->, inner sep=0pt] (w) -- (u);
\end{tikzpicture}
}
\caption{Perturbation on a matching a pair $\langle u, v \rangle$..}
\label{fig:trans1}
\end{figure}

\myparagraph{Open triangles}
A \emph{left open triangle} for $M(\langle u,v \rangle) = y$ is a triple $t = \langle u, v, w \rangle$ with $w \in U$ and $M(\langle w, v \rangle) = \overline{y}$.
%
%
In such a \emph{left open triangle}, $u$ $v$, and $w$  are dubbed the \emph{free} record,  \emph{pivot} record, and the \emph{support} record, respectively. 
Analogously, we can define a \emph{right open triangle}, with the support record from the $V$ table.
%
%
%
For the sake of simplicity, going forward we mostly refer to \emph{left open triangle} cases. All definitions and methods apply to \emph{right open triangles} analogously.

\myparagraph{Open triangle perturbations} 
%
Given a left open triangle, we generate a \emph{perturbed copy} $u^{\prime}$ of the free record $u$ from the support record $w$ by means of a \emph{perturbing record function} $\psi(u, w, A)$, with $A \subseteq A_U$. The perturbing function generates $u^{\prime}$ by replacing sequences of tokens of all the attributes in $A$ in the free record $u$ with their corresponding sequences of tokens from the support record $w$, i.e., $ u^{\prime}[a] \leftarrow w[a], \forall a \in A$. 

In right open triangles, where $v \in V$ is the free record and $u \in U$ is the pivot record, we select $w \in V$ and then build perturbed copies of $v$ by replacing sequences of tokens of attributes in $A_V$.

\ourmethod{} builds perturbed copies in a data-driven way, using sequences of tokens that come from the training set distribution and hence are more likely to be correctly classified by the ER system. 
Perturbed copies are used to compute saliency and counterfactual explanations according to the probabilistic framework developed in~\cite{watson2021local}, which associates the former to the probability of necessity, and the latter to the probability of sufficiency.



\subsection{Saliency Explanations}



We define the \emph{saliency} of an attribute $a \in A_U$ (resp. in $A_V$) in the prediction outcome $M(\langle u, v \rangle) = y$ as the probability that changing the value of $a$ in $u$ (resp. $v$) is a \emph{necessary} factor for flipping the outcome of the prediction.

To compute such a probability, if $a \in A_U$, we rely on a set $W$ of support records for the free node $u$: $W = \{ w | w \in U, M(\langle w,\ v \rangle) = \overline{y} \}$, each record corresponding to a left open triangle $\langle u,v,w \rangle$. Otherwise, if $a \in A_V$, we rely on right open triangles analogously. In the following, for sake of simplicity, we focus on the former case.

Let $\mathcal{U}_{w, a}$ denote the set of perturbed copies of $u$ generated by a support record $w$ by changing all the possible sets of attributes $A \subseteq A_U$ that includes a given attribute $a$. 

\begin{equation*}
    \mathcal{U}_{w, a} = \{\psi(u, w, A) | A \in \mathcal{P}(A_U), a \in A \}
\end{equation*}

\noindent
where  $\mathcal{P}(A_U)$ is the powerset of $A_U$. Let $\mathcal{U}_{a} = \bigcup_{w \in W} \mathcal{U}_{w,a}$. 



\begin{example}
Consider the records in Figure~\ref{fig:example}. Suppose we want to produce an explanation of the Ditto prediction $M(u_1,v_1)= \mathbb{T}$. A left triangle that uses $u_2$ as a support record (assuming  $M(u_2, v_1)=\mathbb{F}$) creates $4$ perturbed copies of $u_1$:
\begin{equation*}
\begin{split}
\mathcal{U}^{\prime} & _{u_2,Name_{Abt}} = \{  
\psi(u_1, u_2, \{Name_{Abt}\}),\\
&\psi(u_1, u_2, \{Name_{Abt}, Description_{Abt}\}),\\
&\psi(u_1, u_2, \{Name_{Abt}, Price_{Abt}\}),\\
&\psi(u_1, u_2, \{Name_{Abt}, Description_{Abt}, Price_{Abt}\}) \}
\end{split}
\end{equation*}

\noindent
For the sake of simplicity, we show here only $2$ of such perturbed copies (copied values are in boldface):


\begin{myitemize}
    \item $\psi(u_1, u_2, \{Name_{Abt}\}) =$\\ $\langle \textbf{``altec lansing inmotion portable audio system ...''}$, \\ $``sony\ bravia\ theater\ black micro...''$,  $NaN\rangle$
    \item $\psi(u_1, u_2, \{Name_{Abt}, Description_{Abt}\}) =$ \\
    $ \langle \textbf{``altec lansing inmotion portable audio system ...''}$,\\ $\textbf{``altec lansing inmotion ipod portable audio system}$ \\  $\textbf{im600usb...''}, NaN\rangle$
\end{myitemize}
\end{example}


\myparagraph{Saliency score} Given a prediction to explain $M(\langle u, v \rangle) = y$ the \emph{saliency score} of an attribute $a \in A_U$, denoted as $\phi_{a}$, corresponds to the probability that the value of $a$ is changed with values coming from any $w \in W$, conditioned on the fact that $M(\langle w, v \rangle)$ flips the prediction, formally:

\begin{equation}
    \phi_{a} = P(u' \in \mathcal{U}_{a} | M(\langle u^{\prime}, v\rangle) = \overline{y})
\label{eq:saliency}
\end{equation}
\noindent


The saliency score for the attributes belonging to the schema of $A_U$ is $\Phi_{A_U} = \{\phi_{a_{U_1}}, \dots , \phi_{a_{U_h}}\}$. The saliency score for the attributes belonging to $A_V$ (i.e., for the schema of the right attribute $v$ of the input pair of the prediction) are computed accordingly.
Finally, a saliency explanation for an ER prediction $M(\langle u, v \rangle) = y$ is composed by the saliency scores for all the attributes in $A_U \cup A_V$, $\Phi = \Phi_{A_U} \cup \Phi_{A_V}$.  



\subsection{Counterfactual explanations}


\emph{Counterfactual} explanations are associated with the concept of sufficiency. That is, the probability that changing the value of a certain set of attributes is a \emph{sufficient} factor for flipping the outcome of a prediction.

Let $\mathcal{U}_{A}$ be the set of perturbed copies $u'$ altered by changing all the attributes in $A \subset A_{U}$, using a set of support records $W$ from left open triangles.

\begin{equation*}
    \mathcal{U}_{A} = \{ \psi(u, w, A) | w \in W \}
\end{equation*}

The probability of sufficiency that changing a given set of attributes $A \subset A_U$ in the original pair $\langle u, v \rangle$ results in flipping the prediction from $y$ to $\overline{y}$ corresponds to the probability that $M(\langle u, v \rangle)$ is flipped conditioned on the fact that the attributes $A$ have been changed in record $u$.

\begin{equation}
    \chi_{A} = P(M(\langle u', v\rangle) = \overline{y} | u' \in \mathcal{U}_{A}) 
\label{eq:sufficiency}
\end{equation}

For each $A$ such that $\chi_{A} > 0$ we can generate a counterfactual explanation as we have at least one $\langle u^{\prime}, v \rangle$ such that $M(\langle u^{\prime}, v\rangle) = \overline{y}$ and $u^{\prime} = \psi(u, w, A)$ for a given $w$.

We define a counterfactual explanation for $M(\langle u,v \rangle) = y$ as a pair of records $\langle u^{\prime}, v \rangle$ whose changed attributes $A \subset A_U$ have the highest probability of sufficiency that changing them yields a prediction flip, with $A$ being as small as possible.

\begin{equation}
    A^{\star} = \operatorname*{argmin}_{A} (|\operatorname*{argmax}_{A \subset \mathcal{P}(A_U) \setminus A_U) } \chi_{A}|)
    \label{eq:astar}
\end{equation}

Symmetrically we can find counterfactual explanations on the attributes in $A_V$ using right open triangles.

Note that, while providing a counterfactual explanation in terms of a proper example, \ourmethod{} also provides a human interpretable measure of the importance of the example. The value $\chi_{A^{\star}}$ associated with the set of attributes $A^{\star}$ reveals that by changing all the attributes in $A^{\star}$ the original predicted outcome flips with a probability of $\chi_{A^{\star}}$.

\subsection{Obtaining triangles}
\label{sec:obtaintri}

Support records from open triangles are used to change the sequences of tokens contained in attributes in the free record of a prediction to be explained. 
Computing the scores $\phi$ and $\chi$ defined in Equations~\ref{eq:saliency} and~\ref{eq:sufficiency} require calculating how frequently such attribute modifications co-occur with a flipped outcome. 
Therefore \ourmethod{} needs an equal number of left and right open triangles to be generated to explain each prediction.

Left open triangles for a prediction $\langle u, v \rangle$ are obtained by calling the classifier $M$ on all the records $w \in U \setminus \{u\}$ such that  $M(\langle w, v\rangle)=\overline{y}$.
Symmetrically, right open triangles for a prediction $\langle u, v \rangle$ are obtained by calling the classifier $M$ on all the records $q \in V \setminus \{v\}$ such that  $M(\langle u, q\rangle)=\overline{y}$.

In case the number of open triangles generated this way is smaller than expected, \ourmethod{} adopts a simple data augmentation scheme to generate more record pairs to evaluate, defined as follows.
The value of an attribute $a_{U_i}$ in a record $w$ is a sequence of tokens (strings separated by white space) $w[a_{U_i}] = \{s_1, s_2, ..., s_n\}$.
%
For each record $w$ in $U$, we generate a new set of records $W_w$, by changing each possible combination of attributes in $w$ by dropping the first-k or the last-k tokens, with $k$ varying between $1$ and $n-1$.

Intuitively larger numbers of triangles are desirable, in order to more accurately approximate the probability values for  necessity and sufficiency.
An experimental evaluation of the impact of the number of triangles used to generate explanations is provided in Section~\ref{sec:trimpact}.



\section{Computing Necessity  and  Sufficiency Probabilities}
\label{sec:lattices}

In order to calculate the probability of necessity of an attribute ($\phi_a$ with $a \in A_U$ or $a \in  A_V$), which provides us its saliency score, and the probability of sufficiency of a set of attributes ($\chi_A$ with $A \subset A_U$ or $A \subset A_V$), which allows us to obtain a counterfactual explanation, we use a frequentist approach. Namely we count: 
\begin{myitemize}
    \item the number of times an attribute is changed with respect to the number of actual flips (eq.~\ref{eq:saliency});
    \item the number of times changing a set of attributes results in a flip, with respect the number of times that the set of attributes is changed (eq.~\ref{eq:sufficiency}).
\end{myitemize}

Computing the above numbers exactly would require to process multiple open triangles and test all the corresponding perturbed copies (namely $|W| \cdot ((|\mathcal{P}(A_U)|-2) + (|\mathcal{P}(A_V)|-2))$ copies)%
\footnote{We do not need to compute the empty set and the entire set of attributes $A_U$ and $A_V$.} of the free record and, for each of them, computing the prediction. We can, however, be more efficient by \emph{inferring} which attributes result in a flip, as described in the following. 

Given a prediction $M(\langle u, v \rangle)=y$, for each left (resp. right) triangle $t=\langle u, v, w \rangle$, with $w \in W$, we build a lattice on the partial order between the elements of the power set $\mathcal{P}(A_U)$ (resp. $\mathcal{P}(A_V)$) and the subset inclusion relation.  
Figure~\ref{fig:lattice-single} shows a lattice structure for the power set of the attributes of the \emph{Abt} schema of Figure~\ref{fig:example} (for now, ignore the colors of the edges and the $\gamma()$ function).  

\begin{figure}
\centering
\resizebox{0.4\textwidth}{!}{
\begin{tikzpicture}
  \node (max) at (0,4) {$\scriptstyle\gamma(\{N_{Abt},D_{Abt},P_{Abt}\}) = 1$};
  \node (a) at (-3,2) {$\scriptstyle\gamma(\{N_{Abt}, D_{Abt}\}) = 1$};
  \node (b) at (0,2) {$\scriptstyle\gamma(\{N_{Abt}, P_{Abt}\}) = 1$};
  \node (c) at (3,2) {$\scriptstyle\gamma(\{D_{Abt}, P_{Abt}\} = 1$};
  \node (d) at (-3,0) {\textcolor{red}{$\scriptstyle\gamma(\{N_{Abt}\} = 1$}};
  \node (e) at (0,0) {\textcolor{red}{$\scriptstyle\gamma(\{D_{Abt}\}) = 1$}};
  \node (f) at (3,0) {$\scriptstyle\gamma(\{P_{Abt}\} = 0$};
  \node (min) at (0,-2) {$\scriptstyle\gamma(\emptyset) ) = 0$};
  \draw (min) -- (d) -- (a) -- (max) -- (b) -- (f)
  (e) -- (min) -- (f) -- (c) -- (max)
  (d) -- (b);
  \draw[preaction={draw=white, -,line width=6pt}] (a) -- (e) -- (c);
  \draw[red, very thick] (d) -- (e);
  \draw[blue, ultra thick] (d) -- (a) -- (max);
  \draw[blue, ultra thick] (d) -- (b) -- (max);
  \draw[blue, ultra thick] (e) -- (a);
  \draw[blue, ultra thick] (e) -- (c) -- (max);
  
\end{tikzpicture}
}
\caption{A lattice structure for a left open triangle on a pair $\langle u_i, v_i \rangle$ from \emph{Abt-Buy} dataset. Nodes are tagged by \emph{flipping} operator $\gamma(\cdot)$. For the sake of readability, we have abbreviated the attribute names with their initials. A minimal flipping antichain $\{\{N_U\}, \{T_U\}\}$ is highlighted in red.}
\label{fig:lattice-single}
\end{figure}

Then, we tag each node $A$ of the lattice with $\gamma(A)$, where:

\begin{equation*}
    \gamma(A) = \mathbbm{1}(M(\langle u',v \rangle) = \overline{y})
\end{equation*}
with $u' = \psi(u, w, A)$. 

Essentially, each node is tagged $1$ if copying the values of attributes in $A$ from the support record $w$ into the corresponding attributes of the free record $u$ leads to flipping the original output $y$, $0$ otherwise.





Continuing our example, let us suppose that any subset of $\mathcal{P}(\{Name_{Abt}, Description_{Abt}, Price_{Abt}\})$ except $\{Price_{Abt}\}$ flips the prediction $M(\langle u, v \rangle)$,
Figure~\ref{fig:lattice-single} shows the lattice structure of our running examples with the nodes tagged accordingly.

Inspired by the work in~\cite{tao2018entity}, we can make the simplifying assumption that the classifier $M$ is \emph{monotone}: if copying the values of the attributes in $A$ from the support record $w$ to the free record $u$ yields a flipped outcome, then we expect that copying values from a superset $A^{\prime} \supset A$ the same way will also flip the prediction. Formally, $\gamma(A) = 1 \implies \gamma(A^{\prime}) = 1, \forall A^{\prime} \supset A$.
An empirical evaluation of the veracity of this property is provided in Section~\ref{sec:monotoneeval}.

Consider Figure~\ref{fig:lattice-single}: assuming $M$ is monotone, if perturbing $u$ copying only $\{Name_{Abt}\}$ flips the prediction, then also all the perturbations built using supersets of $\{Name_{Abt}\}$%
\footnote{Namely: $\{Name_{Abt}, Description_{Abt}\}$, $\{Name_{Abt}, Price_{Abt}\}$, $\{Name_{Abt}, Description_{Abt}, Price_{Abt}\}$.} will flip the predictions, and thus we do not need to compute them.  




Given a lattice $L$, an \emph{antichain} is a set of nodes in $L$ that are not pairwise comparable according to the partial order relations of the lattice. We define the concept of \emph{flipping antichain} as a lattice antichain formed by nodes tagged with $1$ (that is, nodes for which the prediction flipped).  





Given a set of flipping antichains $\Gamma$, a flipping antichain $\eta \in \Gamma$ is \emph{minimal} (Minimal Flipping Antichain, or MFA in short) if any other flipping antichain in $\Gamma$ only contains elements that are supersets of elements of $\eta$ (i.e., any subset of the attribute sets in $\eta$ do not cause a flip).



\begin{figure}
\centering
\subfigure[\label{lat:a}$w_1$]{
    \resizebox{0.45\columnwidth}{!}{
    \begin{tikzpicture}
      \node (max) at (0,4) {$ N,D,P = 1$};
      \node (a) at (-3,2) {$ N,D = 1$};
      \node (b) at (0,2) {$ N,P = 1$};
      \node (c) at (3,2) {$ D,P = 1$};
      \node (d) at (-3,0) {\textcolor{red}{$ N = 1$}};
      \node (e) at (0,0) {\textcolor{red}{$ D = 1$}};
      \node (f) at (3,0) {$ P = 0$};
      \node (min) at (0,-2) {$ \emptyset = 0$};
      \draw (min) -- (d) -- (a) -- (max) -- (b) -- (f)
      (e) -- (min) -- (f) -- (c) -- (max)
      (d) -- (b);
      \draw[preaction={draw=white, -,line width=6pt}] (a) -- (e) -- (c);
      \draw[red, very thick] (d) -- (e);
      \draw[blue, ultra thick] (d) -- (a) -- (max);
      \draw[blue, ultra thick] (d) -- (b) -- (max);
      \draw[blue, ultra thick] (e) -- (a);
      \draw[blue, ultra thick] (e) -- (c) -- (max);
      
    \end{tikzpicture}
    }
}
\subfigure[\label{lat:b}$w_2$]{
    \resizebox{0.45\columnwidth}{!}{
    \begin{tikzpicture}[]
      \node (max) at (0,4) {$ N,D,P = 1$};
      \node (a) at (-3,2) {$ N,D = 1$};
      \node (b) at (0,2) {$ N,P = 1$};
      \node (c) at (3,2) {\textcolor{red}{$ D,P = 1$}};
      \node (d) at (-3,0) {\textcolor{red}{$ N = 1$}};
      \node (e) at (0,0) {$ D = 0$};
      \node (f) at (3,0) {$ P = 0$};
      \node (min) at (0,-2) {$ \emptyset = 0$};
      \draw (min) -- (d) -- (a) -- (max) -- (b) -- (f)
      (e) -- (min) -- (f) -- (c) -- (max)
      (d) -- (b);
      \draw[preaction={draw=white, -,line width=6pt}] (a) -- (e) -- (c);
      \draw[red, very thick] (d) -- (c);
      \draw[blue, ultra thick] (d) -- (a) -- (max);
      \draw[blue, ultra thick] (c) -- (max);
      \draw[blue, ultra thick] (d) -- (b) -- (max);
      
    \end{tikzpicture}
    }
}

\subfigure[\label{lat:c}$w_3$]{
    \resizebox{0.45\columnwidth}{!}{
    \begin{tikzpicture}[]
      \node (max) at (0,4) {$ N,D,P = 1$};
      \node (a) at (-3,2) {$ N,D = 1$};
      \node (b) at (0,2) {$ N,P = 1$};
      \node (c) at (3,2) {$ D,P = 0$};
      \node (d) at (-3,0) {\textcolor{red}{$ N = 1$}};
      \node (e) at (0,0) {$ D = 0$};
      \node (f) at (3,0) {$ P = 0$};
      \node (min) at (0,-2) {$ \emptyset = 0$};
      \draw (min) -- (d) -- (a) -- (max) -- (b) -- (f)
      (e) -- (min) -- (f) -- (c) -- (max)
      (d) -- (b);
      \draw[preaction={draw=white, -,line width=6pt}] (a) -- (e) -- (c);
      \draw[red, very thick] (e) -- (e);
      \draw[blue, ultra thick] (d) -- (a) -- (max);
      \draw[blue, ultra thick] (d) -- (b) -- (max);
      
    \end{tikzpicture}
    }
}
\subfigure[\label{lat:d}$w_4$]{
    \resizebox{0.45\columnwidth}{!}{
    \begin{tikzpicture}[]
      \node (max) at (0,4) {$ N,D,P = 1$};
      \node (a) at (-3,2) {\textcolor{red}{$ N,D = 1$}};
      \node (b) at (0,2) {\textcolor{red}{$ N,P = 1$}};
      \node (c) at (3,2) {\textcolor{red}{$ D,P = 1$}};
      \node (d) at (-3,0) {$ N = 0$};
      \node (e) at (0,0) {$ D = 0$};
      \node (f) at (3,0) {$ P = 0$};
      \node (min) at (0,-2) {$ \emptyset = 0$};
      \draw (min) -- (d) -- (a) -- (max) -- (b) -- (f)
      (e) -- (min) -- (f) -- (c) -- (max)
      (d) -- (b);
      \draw[preaction={draw=white, -,line width=6pt}] (a) -- (e) -- (c);
      \draw[red, very thick] (a) -- (b) -- (c);
      \draw[blue, ultra thick] (a) -- (max);
      \draw[blue, ultra thick] (b) -- (max);
      \draw[blue, ultra thick] (c) -- (max);
      
    \end{tikzpicture}
    }
}

\caption{Example lattice structures.}
\label{fig:lattice3}
\end{figure}

For this reason, when the monotone classification property is satisfied, identifying an MFA $\eta$ saves us from calculating all the predictions corresponding to the perturbations involving supersets of elements in $\eta$.

Assuming monotone classification, performing as few predictions as possible on a lattice $L$ corresponds to finding the largest MFAs in $L$. To this end, we visit the lattice bottom-up with a breadth-first strategy until all the lattice nodes are tagged. For each visited node, we compute the prediction associated to the perturbation corresponding to the attributes of the node. Whenever the prediction flips with respect to the input prediction, we propagate the predicted outcome to all the upward chains leading to the supremum of the lattice.  

\noindent\textbf{Example.} Consider the pair of records $\langle u_1, v_1 \rangle$ in Figure~\ref{fig:example} with $M=$~Ditto and let us focus on explanations for attributes of $u_1$. As $M(\langle u_1, v_1 \rangle)=\mathbb{T}$, we need to identify records $w \in U$ s.t. $M(\langle w, v_1 \rangle)=\mathbb{F}$. Such records, let them be the fictitious records $W=\{w_1, w_2, w_3, w_4\}$, are used as support records for building four left open triangles $\langle u_1, v_1, w \rangle$, $w \in W$, with $u_1$ as the free record and $v_1$ as the pivot. Let the lattices corresponding to the four triangles be those shown in Figure~\ref{fig:lattice3}. Note that all the triangles are left and thus all the lattices' nodes represent subsets of attributes in $A_U$. For sake of brevity, we show only each attribute's initial (i.e., $N$ for $Name_{Abt}$, $D$ for $Description_{Abt}$ and $P$ for $Price_{Abt}$) and omit the $\gamma$ notation. The nodes included in the largest MFA and the edges representing upward paths with flip propagation are highlighted respectively in red and blue. 

When processing $w_1$ (i.e., the open triangle $\langle u_1, v_1, w_1 \rangle$), we get a flip for $\{N\}$ and $\{D\}$ and a non-flip for $\{P\}$. That is, $M(\langle \psi(u_1, w, A), v_1\rangle) = \mathbb{F}$, for $A=\{N\}$ and $A=\{D\}$, while $M(\langle \psi(u_1, w, A), v_1\rangle) = \mathbb{T}$ for $A=\{P\}$. Assuming that $M$ is monotone, we can infer the flip/non-flip results for all the upward nodes in the lattice in Figure~\ref{lat:a} and identify $\{\{N\}, \{D\}$\} as the largest MFA without further testing. 

When processing $w_2$ and $w3$, we get a flip for $\{N\}$ and a non-flip for the other singleton nodes. In those cases, we can infer only $\{N,D\}$ and $\{N,P\}$ while we need to test $\{D,P\}$ explicitly. That is, we need to collect the result of $M(\langle \psi(u_1, w, \{D,P\}), v_1\rangle)$. In the case of $w_2$, the collected result is negative, yielding a flip, and thus we identify $\{\{N\}, \{D,P\}\}$ as the largest MFA (Figure~\ref{lat:b}). In the case of $w_3$, the collected result is positive, yielding a non-flip, and thus the largest MFA consists solely of $\{N\}$ (Figure~\ref{lat:c}).

Finally, when processing $w_4$, we get all non-flips at the first level, meaning that copying only one attribute from $w_4$ is not enough for flipping the prediction. In such a case, we need to test all the attribute pairs explicitly, by collecting the result of $M(\langle \psi(u_1, w, A), v_1\rangle)$, for all $|A|=2, A \in \mathcal{P}(A_U)$. As shown in Figure~\ref{lat:d}, we get all flips, and thus identify $\{\{N,D\}, \{N,P\}, \{D,P\}\}$ as the largest MFA.

In order to compute explanation scores $\phi$ and $\chi$ as in Equations~\ref{eq:saliency} and~\ref{eq:sufficiency} respectively, we need to consider all the nodes corresponding to flips, either tested or inferred. Specifically, in Figures~\ref{lat:a}--\ref{lat:d} we have a total of 19 flips. As for the saliency explanations, we obtain $\phi_{N}=\frac{15}{19}$, $\phi_{D}=\frac{13}{19}$ and $\phi_{P}=\frac{11}{19}$. As for the counterfactual explanations, we get $\chi_{\{N\}}=\frac{3}{4}$ ($4$ is the size of $W$), $\chi_{\{D\}}=\frac{1}{4}$, $\chi_{\{P\}}=0$, $\chi_{\{N,D\}}=1$, $\chi_{\{N,P\}}=1$ and $\chi_{\{D,P\}}=\frac{3}{4}$. Since for this example we have $\max_{A \subset A_U} \chi_{A} = 1$ and $A^*=\{N,D\}$ or $A^*=\{N,P\}$ (note that $A^*$ cannot be $\{N,D,P\}$ in Equation~\ref{eq:astar}). The resulting counterfactual explanations are all the pairs $\langle u', v_1\rangle$ such that $u' \in \{\psi(u,w,\{N,D\} | w \in W\} \cup \{\psi(u,w,\{N,P\} | w \in W\}$, as they all yield a flip. 

\noindent\textbf{The \ourmethod{} algorithm.} Overall, the \ourmethod{} approach is summarized in Algorithm~\ref{alg:certa}. \ourmethod{} keeps counters for sufficiency of sets of attributes ($S$), necessity of an attribute ($N$), and number of flips ($f$). First, it fetches $\tau$ open triangles (line 8); the method \emph{get\_triangles()} generates $\frac{\tau}{2}$ left open triangles from records $w \in U$ and $\frac{\tau}{2}$ right open triangles using records $q \in V$. Then, for each triangle \ourmethod{} builds the corresponding lattice (line 10) and finds the largest minimal flipping antichain (line 11). From the antichain $\eta_{min}$ it derives all the inputs $c$ that flip the prediction, associated to their corresponding set of changed attributes $A$ (line 12) and updates candidate counterfactuals set $C$ with $c$ (line 13), flip counts for $A$ (line 14) and aggregate flip counts $f$ (line 15). Then, for each attribute $a \in A$ it updates the necessity counts (line 17).
\ourmethod{} generates saliency scores $\Phi$ by dividing the necessity counts by the aggregate flip counts (line 19). For counterfactuals, it generates the sufficiency for attribute sets (line 14) and checks whether it is bigger than current maximum sufficiency (line 24) or equal but involving fewer attributes (line 27). This way the golden set of attributes is identified. Finally, it generates the list of counterfactual explanations whose changed attributes correspond to such a golden set (lines 30-33).

\SetKwProg{Fn}{procedure}{:}{} 
\SetKwFunction{FMain}{\ourmethod{}}
\begin{algorithm}[t]
\scriptsize
\SetAlgoLined
\SetKwInOut{Input}{input}\SetKwInOut{Output}{output}
\Input{$M(\langle u,v \rangle) = y$, number of triangles $\tau$, $U$, $V$}
\Output{attributes saliency $\Phi$, set of counterfactual examples $E$}
\Fn{\FMain{$u, v, \tau, M, U, V$}}{
  \ForEach{$A \in \mathcal{P}(A_U) \setminus A_U \cup \mathcal{P}(A_V) \setminus A_V$}{
    $S[A] \leftarrow 0$\;
  }
  \ForEach{$a \in A_U \cup A_V$}{
    $N[a] \leftarrow 0$\;
  }
  $f = 0$\;
  $C \leftarrow \emptyset$\;
  $T \leftarrow get\_triangles(M, u, v, y, U, V, \tau)$\;
  \ForEach{$t \in T$}{
    $L_t = build\_lattice(t)$\;
    $\eta_{min} = get\_lmfa(L_t, M, u, v, y$)\;
    \ForEach{$(c, A) \in  get\_flipped(\eta_{min})$}{
        $C \leftarrow C \cup \{ (c, A) \}$\;
        $S[A] \leftarrow S[A] + 1$\;
        $f \leftarrow f + 1$\;
        \ForEach{$A \in a$}{
            $N[a] \leftarrow 1$\;
        }
    } 
  }
  \ForEach{$a \in A_U \cup A_V$}{
    $\phi_a \leftarrow \frac{N[a]}{f}$\;
    $\Phi \leftarrow \Phi \cup \{\phi_a\}$\;
  }
  $A^\star \leftarrow \emptyset$\;
  $\chi^\star \leftarrow 0$\;
  \ForEach{$A \in \mathcal{P}(A_U) \setminus A_U \cup \mathcal{P}(A_V) \setminus A_V$}{
    \If{$\frac{S[A]}{|T|} > \chi^\star) $}{
       $\chi^\star \leftarrow \frac{S[A]}{|T|}$\;
       $A^\star \leftarrow A$\;
    }
    \ElseIf{$\frac{S[A]}{|T|} == \chi^\star \textbf{and} |A| < |A^\star|$}{
        $\chi^\star \leftarrow \frac{S[A]}{|T|}$\;
        $A^\star \leftarrow A$\;
    }
  }
  $E \leftarrow \emptyset $\;
  \ForEach{$(c,A) \in C$}{
      \If{$A^\star == A$}{
          $E \leftarrow E \cup \{ c \}$\;
      }
  }
 \Return $\Phi, E$\;
 }
\caption{The \ourmethod{} algorithm.}
\label{alg:certa}
\end{algorithm}

\section{Experiments}
\label{sec:exp}


\subsection{Experimental setup}
\label{sec:setup}
We aim to quantitatively measure how explanations generated by CERTA and baselines are effective.
Different quantitative measures of effectiveness exist, depending on the specific type of explanation to evaluate (see Section~\ref{sec:exp-method}).
We seek not to evaluate plausibility via any user study though, as any possible correlation between plausibility and model performance would increase user performance too and thus invalidate any subsequent result \cite{jacovi2020towards}.

We perform separate experiments for saliency and counterfactual explanations, considering appropriate baseline methods respectively.

\myparagraph{Affected models}
We evaluate \ourmethod{} using three recent state-of-the-art ER systems based on deep learning (DL), namely:
\begin{myitemize}
    \item the LSTM model of DeepER~\cite{ebraheem2018distributed}, a DL architecture for ER based on distributed representation of records;
    \item the Hybrid model of DeepMatcher~\cite{mudgal2018deep}, a DL framework based on distributed representation of attributes
    \item the DistilBERT~\cite{DBLP:journals/corr/abs-1910-01108} based model of {\ditto}; \cite{DBLP:journals/pvldb/0001LSDT20}, a DL solution based on the Transformers architecture, with data augmentation and injection of domain knowledge.
\end{myitemize}

\myparagraph{Datasets}
We use the datasets of the DeepMatcher repository,\footnote{\url{https://github.com/anhaidgroup/deepmatcher/blob/master/Datasets.md}} which have been adopted by the above systems for their experimental evaluation.\footnote{We have excluded the Company dataset as it has only one attribute.}
Table~\ref{tab:benchdata} summarizes the main characteristics of each dataset: column ``Matches'' reports the number of matching pairs of the ground truth; ``Records'' and ``Values'' lists the number of records and the number of distinct values in the two sources, respectively.
Each dataset comes with its own test and training set, which we use for training the DL models.

\begin{table}[t]
    \setlength{\tabcolsep}{0.42em}
    \centering
    \scriptsize
    \begin{tabular}{|l|c|c|c|c|}
\hline
\textbf{Dataset} & \textbf{Matches} & \textbf{Attr.s} & \textbf{Records} & \textbf{Values} \\
\hline
\hline
AB (Abt-Buy)  & 5743 & 3 & 1081 - 1092 & 776 - 721 \\
\hline
AG (Amazon-Google)  & 1167 & 3 & 1363 - 3226 & 650 - 1511 \\
\hline
BA (beerAdvo-RateBeer)  & 68 & 4 & 4345 - 3000 & 1807 - 1323 \\ 
\hline
DA (DBLP-ACM) & 2220 & 4 & 2614 - 2292 & 1209 - 1060 \\ 
\hline
DS (DBLP-Scholar) & 5547 & 4 & 2614 - 64263 & 1152 - 32664 \\
\hline
FZ (Fodors-Zagats)  & 110 & 6 & 533 - 331 & 360 - 236 \\
\hline
IA (iTunes-Amazon) & 132 & 8 & 6907 - 55923 & 903 - 6444 \\ 
\hline
WA (Walmart-Amazon) & 962 & 5 & 2554 - 22074 & 1370 - 9504 \\
\hline
DDA (Dirty DBLP-ACM) & 7418 & 4 & 2614 - 2292 & 938 - 840 \\
\hline
DDS (Dirty DBLP-Scholar) & 17223 & 4 & 2614 - 64263 & 909 - 25096 \\
\hline
DIA (Dirty iTunes-Amazon) & 321 & 8 & 6907 - 55923 & 1244 - 6364\\
\hline
DWA (Dirty Walmart-Amazon) & 6144 & 5 & 2554 - 22074 & 1001 - 7347 \\
\hline
\end{tabular}
    \caption{Datasets for experimental evaluation.}
    \label{tab:benchdata}
\end{table}

\subsection{Baseline methods} 
For conducting quantitative evaluations of the effectiveness of \ourmethod{}, we identify two sets of baselines, one  of saliency explanations, and one for counterfactual explanations.

\myparagraph{Saliency method baselines}
We compare the saliency explanations generated by \ourmethod{} both with methods that are aware of semantics of the ER task, and with methods that agnostic with respect to the semantics of the classification task.
For ER semantics aware saliency explanation methods, we compare against Mojito~\cite{di2019interpreting} (which is based on LIME~\cite{ribeiro2016should}) and LandMark~\cite{DBLP:conf/edbt/BaraldiBP021}. For Mojito we use the \textit{mojito-drop} technique for explaining \emph{Match} predictions and the \textit{mojito-copy} technique for explaining \emph{Non-Match} predictions, in line with the semantics of the method. 

For task agnostic methods, we use SHAP~\cite{DBLP:conf/nips/LundbergL17} within our evaluation as it is one of the most popular black box explanation methods.

\myparagraph{Counterfactual method baselines}
Also for the counterfactual explanations, we compare the results generated by \ourmethod{} with both semantics aware and semantics agnostic counterfactual methods. 
As semantics agnostic baseline, we compare against DiCE~\cite{DBLP:conf/fat/MothilalST20}, a black box counterfactual explanation generation method.
To the best of our knowledge, no ER specific counterfactual framemwork exists yet, therefore we adapt the \emph{LIME-C} and \emph{SHAP-C} counterfactual expanation methods~\cite{DBLP:journals/adac/RamonMPE20} to work within the ER setting, as follows:
\begin{myitemize}
    \item we treat input record pairs as text;
    \item for LIME-C we adopt \emph{Mojito} instead of plain LIME, to have a better fit with the ER setting.
\end{myitemize}



\subsection{Evaluation methodology}
\label{sec:exp-method}
We consider different metrics for evaluating different kinds of explanations.
Note that for each dataset, all the evaluated metrics are computed on all the examples contained in the corresponding test set. For \ourmethod{} we use $\tau = 100$ triangles in all our experiments, unless specified. 
In Section~\ref{sec:trimpact} we present experiments that show the robustness of \ourmethod{} with respect to this parameter.

For saliency explanations we use the quantitative explanation evaluation metrics of \emph{Faithfulness} and \emph{Confidence indication}~\cite{DBLP:conf/emnlp/AtanasovaSLA20}. 
\begin{myitemize}
\item \emph{Faithfulness} aims at detecting whether attributes that are important according to an explanation are actually important to the ER system. Intuitively, modifying attributes with a high saliency should cause a significant change in the score of the prediction,  while changing poorly salient attributes should not alter the prediction much. Faithfulness measures the area under the threshold-performance curve (AUC). Thresholds indicate the fraction of attributes that have to be masked. The attributes to be masked are taken from the saliency explanation, in descending saliency score order. The set of thresholds used is $\{0.1, 0.2, 0.33, 0.5, 0.7, 0.9\}$ and the performance measure is the F1 of the model $M$. Faithful explanations are expected to induce a higher F1 drop as more salient attributes are incrementally masked. Low AUC values indicate high faithfulness.
\item
\emph{Confidence indication} seeks to find out whether an explanation is a good \emph{proxy} of the confidence of the system, e.g., generally low saliency scores should correspond to predictions the system has low confidence on, while in presence of highly salient attributes the system should be highly confident about the prediction.
Confidence indication is calculated as the \emph{mean absolute error} (MAE) of a logistic regression classifier trained with saliency explanation scores for match/nomatch (input) and the actual score of the model (label). A low MAE value indicates that the model’s score can be easily identified by looking at the produced explanations.
\end{myitemize}

Quality of counterfactual explanations are evaluated by means of the \emph{Proximity}, \emph{Sparsity} and \emph{Diversity} metrics~\cite{DBLP:conf/fat/MothilalST20}.%
\footnote{Another metric defined in ~\cite{DBLP:conf/fat/MothilalST20} is \emph{Validity}, which measures the fraction of examples returned by a method that are actually counterfactuals, that is, that flip the prediction. However, \ourmethod{} produces by construction counterfactual explanations, while DiCE also returns examples that do not. Then, for a fair comparison, we do not report experimental results based on Validity.}

\begin{myitemize}
\item 
\emph{Proximity} captures how similar a generated counterfactual is to the original input and is calculated as the mean of attribute-wise distances between a counterfactual example and the original input pair. Proximity for a set of examples is simply the average proximity over all the examples.
\item 
\emph{Sparsity} captures the number of changed attributes between the original input and a generated counterfactual.
\item
\emph{Diversity} measures attributes-wise distances between each pair of counterfactual examples to quantify the expressive power of a counterfactual generation algorithm.
\end{myitemize}

For diversity, sparsity and proximity metrics, higher values are better.
To conclude the evaluation of counterfactual explanations, we also report the average number of generated counterfactual explanations by each considered method. 

\begin{table*}[ht]
\centering
\scriptsize
\begin{tabular}{|c||c|c|c|c||c|c|c|c||c|c|c|c|}

\cline{2-13}
\multicolumn{1}{c||}{} & \multicolumn{4}{c||}{DeepER}
& \multicolumn{4}{c||}{DeepMatcher}
& \multicolumn{4}{c|}{{\ditto}}
\\ \hline
\hline
Dataset & \ourmethod{} & LandMark & Mojito & SHAP & \ourmethod{} & LandMark & Mojito & SHAP & \ourmethod{} & LandMark & Mojito & SHAP \\
\hline
\hline
AB& \textbf{0.006} & 0.12 & 0.03 & 21.49 & \textbf{17.51} & 17.56 & 19.59 & 18.21 & \textbf{0.25} & 0.31 & 0.3 & 0.32 \\
AG& \textbf{0.03} & 0.13  & 0.06 & 0.16 & \textbf{1.42} & 5.17  & 4.71 & \textbf{1.42} & \textbf{0.31} & 0.33  & \textbf{0.31} & 0.35 \\
BA& \textbf{0.003}  & 0.23  & 0.17  & 0.21 & \textbf{8.18}  & 25.17  & 27.71  & 9.13 & \textbf{0.24}  & 0.39  & 0.37  & 0.36 \\
DA& \textbf{0.04}  & 0.33  & 0.09  & 0.17 & \textbf{20.23}  & 34.46  & 35.58  & 34.99 & \textbf{0.14}  & 0.15  & \textbf{0.14}  & 0.41 \\
DS& 0.42 & 0.50  & \textbf{0.32} & 0.44 & 34.9 & 26.4  & 52.7 & \textbf{21.59} & \textbf{0.04} & 0.10 & 0.12 & 0.11 \\
FZ& \textbf{0.336} & 0.338  & 0.42 & 0.34 & \textbf{4.46} & 9.75  & 4.71 & 4.71 & 0.23 & 0.39 & 0.41 & \textbf{0.22} \\
IA & \textbf{0.03} & 0.23  & 0.11 & 0.16  & \textbf{25.72} & 41.32  & 46.23 & 41.08  & \textbf{0.67} & 0.69 & 0.68 & 0.68 \\
WA& \textbf{0.02} & 0.25  & 0.38 & 0.09 & \textbf{10.49} & 10.99  & 38.6 & 29.53 & \textbf{0.57} & 0.64 & 0.59 & 0.59 \\
DDA& 0.28 & 0.52  & \textbf{0.26} & 0.44 & \textbf{17.51} & 29.3  & 30.97 & 61.41 & \textbf{0.34} & 0.41 & 0.41 & 0.44 \\
DDS & \textbf{0.45} & 0.48  & 0.46 & 0.49 & \textbf{5.85} & 6.12  & 8.84 & 8.31 & \textbf{0.09} & \textbf{0.09} & 0.12 & 0.46 \\
DIA & \textbf{0.01}  & 0.17   & 0.06  & 0.15  & \textbf{33.66}  & 34.21   & 30.18  & 30.84  & \textbf{0.12} & 0.23 & 0.19 & 0.51 \\
DWA & \textbf{0.04} & 0.05 & 0.05  & 0.23  & \textbf{11.81} & 14.15 & 17.78  & 23.5  & \textbf{0.07} & 0.08 & 0.08 & 0.09 \\
\hline
\end{tabular}
\caption{Faithfulness evaluation on saliency explanations.}
\label{exp:faithfulness}
\end{table*}

\begin{table*}[ht]
\centering
\scriptsize
\begin{tabular}{|c||c|c|c|c||c|c|c|c||c|c|c|c|}

\cline{2-13}
\multicolumn{1}{c||}{} & \multicolumn{4}{c||}{DeepER}
& \multicolumn{4}{c||}{DeepMatcher}
& \multicolumn{4}{c|}{{\ditto}}
\\ \hline
\hline
Dataset & \ourmethod{} & LandMark & Mojito & SHAP & \ourmethod{} & LandMark & Mojito & SHAP & \ourmethod{} & LandMark & Mojito & SHAP \\
\hline
\hline
AB& \textbf{0.021} & 0.026   & 0.025 & 0.023 & \textbf{0.016} & 0.12   & 0.096 & 0.099 & 0.098 & 0.121 & 0.14 & \textbf{0.045} \\
AG& 0.113 & 0.15  & 0.214 & \textbf{0.098} & \textbf{0.015} & 0.101  & 0.048 & 0.021 & 0.01 & 0.01  & 0.01 & 0.01 \\
BA& \textbf{0.02}  & 0.05  & 0.03  & \textbf{0.02} & 0.11  & 0.126  & 0.12  & \textbf{0.10}  & \textbf{0.298}  & 0.326  & 0.474  & 0.376  \\
DA& \textbf{0.182} & 0.32  & 0.221 & 0.663 & \textbf{0.002} & 0.005  & 0.003 & 0.003 & \textbf{0.104} & 0.151  & 0.126 & 0.115 \\
DS& \textbf{0.213} & 0.308  & 0.292 & 0.248 & 0.046 & 0.049  & \textbf{0.018} & 0.032 & \textbf{0.046} & 0.049  & 0.054 & 0.047 \\
FZ& 0.488 & 0.488  & \textbf{0.396} & 1.93 & \textbf{0.002} & 0.103  & 0.009 & 0.055 & \textbf{0.039} & 0.223  & 0.186 & 0.064 \\
IA& \textbf{0.238} & 0.342  & 0.325 & 0.358 & \textbf{0.281} & 0.364  & 0.295 & 0.289 & \textbf{0.071} & 0.094  & 0.129 & 0.13 \\
WA& \textbf{0.041} & 0.081  & 0.089 & 0.053 & \textbf{0.021} & 0.231  & 0.044 & 0.035 & \textbf{0.015} & 0.08  & 0.051 & 0.046 \\
DDA & \textbf{0.375} & 0.247  & 0.356 & 0.252 & \textbf{0.204} & 0.276  & 0.407 & 0.294 & 0.325  & 0.149 & 0.424  & \textbf{0.07}  \\
DDS & \textbf{0.102}  & 0.144   & 0.171  & 0.14  & \textbf{0.08}  & 0.09   & 0.09  & \textbf{0.08} & \textbf{0.062} & 0.133  & 0.128  & 0.127  \\
DIA & 0.225 & \textbf{0.198}  & 0.23  & 0.233  & \textbf{0.047} & 0.054  & 0.05  & 0.08  & \textbf{0.047} & 0.054  & 0.05 & \textbf{0.047}  \\
DWA & \textbf{0.131} & 0.225  & 0.207 & 0.145  & \textbf{0.251} & 0.269  & 0.272 & 0.263 & \textbf{0.204} & 0.276  & 0.407 & 0.294 \\
\hline
\end{tabular}
\caption{Confidence Indication evaluation on saliency explanations.}
\label{exp:confidence}
\end{table*}

\subsection{Results}

\myparagraph{Saliency explanations}
\label{sec:exp-saliency}
In Table~\ref{exp:faithfulness} we report an evaluation of the faithfulness of the saliency explanations generated using \ourmethod{} versus the all identified baselines. 
For the DeepER model \ourmethod{} reports the best faithfulness measure, but for the DS and DDA datasets, where Mojito is the most faithful (\ourmethod{} being the second most faithful).
For DeepMatcher \ourmethod{} reports the best faithfulness measure, but for the DS dataset where SHAP results in being more faithful; there is also a tie between SHAP and \ourmethod{} on the AG dataset.
For the \ditto{} model \ourmethod{} is the most faithful in almost all the cases, SHAP has a slightly better faithfulness measure for the FZ dataset; there are also two ties between \ourmethod{} and Mojito (DA and AG) and one between \ourmethod{} and LandMark (DDS).

In Table~\ref{exp:confidence} we report an evaluation of the confidence indication of the saliency explanations generated using \ourmethod{} versus the all identified baselines.
\ourmethod{} is the most indicative of the confidence of DeepER for most of the datasets, SHAP wins on the AG dataset, Mojito wins on the FZ dataset while LandMark wins on the DIA dataset.
\ourmethod{} is the most indicative of the confidence of DeepMatcher for most of the datasets, two exceptions relate to BA dateset (SHAP wins) and DS (Mojito wins).
Finally, \ourmethod{} is the most indiciative of the confidence for \ditto{} on most of the datasets, SHAP performs better for AB and DDA datasets and ties on DIA.
   
\begin{table*}[!]
\centering
\scriptsize
\begin{tabular}{|c||c|c|c|c||c|c|c|c||c|c|c|c|}

\cline{2-13}
\multicolumn{1}{c||}{} & \multicolumn{4}{c||}{DeepER}
& \multicolumn{4}{c||}{DeepMatcher}
& \multicolumn{4}{c|}{{\ditto}}
\\ \hline
\hline
Dataset & \ourmethod{} & DiCE & SHAP-C & LIME-C & \ourmethod{} & DiCE & SHAP-C & LIME-C & \ourmethod{} & DiCE & SHAP-C & LIME-C \\
\hline
\hline
AB& \textbf{0.74} & 0.72 & 0.35 & 0.42 & \textbf{0.56} & 0.55 & 0.51 & 0.48 & \textbf{0.55} & 0.52 & 0.52 & 0.28 \\
AG& \textbf{0.51} & 0.49  & 0.33 & 0.31 & 0.66 & \textbf{0.72}  & 0.62 & 0.66 & \textbf{0.94} & 0.51  & 0.38 & 0.49 \\
BA& 0.37  & \textbf{0.59}  & 0.35  & 0.41 & 0.3  & \textbf{0.53}  & 0.18  & 0.28 & \textbf{0.37}  & 0.22  & 0.2  & 0.35 \\
DA& \textbf{0.49}  & 0.44  & 0.18  & 0.41 & \textbf{0.58}  & \textbf{0.58}  & 0.41  & 0.52 & \textbf{0.58}  & 0.49  & 0.48  & 0.38 \\
DS& \textbf{0.63} & 0.6  & 0.38 & 0.55 & 0.55 & 0.55  & \textbf{0.62} & 0.52 & \textbf{0.39} & 0.32 & 0.36 & 0.32 \\
FZ& \textbf{0.52} & 0.41  & 0.39 & 0.48 & \textbf{0.63} & 0.49 & 0.53  & 0.48 & \textbf{0.92} & 0.48 & 0.74 & 0.81  \\
IA & 0.59 & \textbf{0.67}  & 0.21 & 0.55  & \textbf{0.52} & 0.25  & 0.36 & 0.43  & 0.14 & 0.09 & 0.04 & \textbf{0.34} \\
WA & 0.41 & \textbf{0.61}  & 0.39 & 0.4 & 0.35 & 0.3  & \textbf{0.39} & \textbf{0.39} & \textbf{0.49} & 0.35 & 0.31 & 0.15 \\
DDA& \textbf{0.67} & 0.66 & 0.57 & 0.59 & \textbf{0.58} & 0.55  & 0.44 & 0.55 & \textbf{0.59} & 0.4 & 0.25 & 0.39 \\
DDS& \textbf{0.45} & 0.41  & 0.25 & 0.39 & \textbf{0.59} & \textbf{0.59} & 0.58 & 0.39 & \textbf{0.34} & 0.41 & 0.41 & 0.44 \\
DIA & \textbf{0.49}  & 0.38   & 0.39  & 0.35  & 0.67  & \textbf{0.72}   & 0.62  & 0.67  & \textbf{0.66} & 0.49 & 0.4 & 0.55 \\
DWA & \textbf{0.52} & 0.51 & 0.38  & 0.49  & \textbf{0.76} & 0.72 & \textbf{0.76}  & 0.62  & \textbf{0.68} & 0.59 & 0.51 & 0.39 \\
\hline
\end{tabular}
\caption{Proximity evaluation on counterfactual explanations.}
\label{exp:prox}
\end{table*}

\begin{table*}[ht]
\centering
\scriptsize
\begin{tabular}{|c||c|c|c|c||c|c|c|c||c|c|c|c|}

\cline{2-13}
\multicolumn{1}{c||}{} & \multicolumn{4}{c||}{DeepER}
& \multicolumn{4}{c||}{DeepMatcher}
& \multicolumn{4}{c|}{{\ditto}}
\\ \hline
\hline
Dataset & \ourmethod{} & DiCE & SHAP-C & LIME-C & \ourmethod{} & DiCE & SHAP-C & LIME-C & \ourmethod{} & DiCE & SHAP-C & LIME-C \\
\hline
\hline
AB& \textbf{0.9} & 0.81 & 0.89 &0.87 & \textbf{0.91} & 0.82& 0.85 & 0.86 & \textbf{0.93} & 0.87 & 0.85 & 0.1 \\
AG&0.88 & 0.87 & 0.9 & \textbf{0.91} & \textbf{0.94} & 0.92 & 0.93 & 0.92 & \textbf{0.88} & 0.87 & 0.78 & 0.63 \\
BA&\textbf{0.89} & 0.83 & \textbf{0.89} & 0.78 & \textbf{0.96} & 0.89 & 0.95 & 0.93 & \textbf{0.96} & 0.95 & 0.24 & 0.92 \\
DA& \textbf{0.96} & 0.81 & 0.95 & 0.88& \textbf{0.94} & 0.88 & 0.91 & 0.9& \textbf{0.92} & 0.89 & 0.71 & 0.7 \\
DS& \textbf{0.91} & 0.81 & 0.89 & 0.89& \textbf{0.98} & 0.93 & 0.92 & 0.89& \textbf{0.91} & \textbf{0.91} & 0.64 & 0.65 \\
FZ& \textbf{0.92} & 0.91 & 0.83 & 0.88& \textbf{0.93} & \textbf{0.93} & 0.77 & 0.92& 0.91 & 0.75 & 0.89 & \textbf{0.93} \\
IA& \textbf{0.93} & 0.92 & 0.84 & 0.9& \textbf{0.99} & 0.97 & 0.96 & 0.95& \textbf{0.99} & \textbf{0.99} & \textbf{0.99} & 0.96 \\
WA& 0.89 & 0.83 & \textbf{0.94} & 0.91& \textbf{0.92} & 0.89 & 0.89 & 0.81& \textbf{0.96} & 0.94 & 0.9 & 0.74 \\
DDA & \textbf{0.91} & 0.85 & 0.87 & 0.84& \textbf{0.94} & 0.78 & \textbf{0.94} & 0.93 & \textbf{0.95} & 0.93 & 0.84 & 0.72 \\
DDS & 0.9 & 0.87 & 0.89 & \textbf{0.91} & \textbf{0.95} & 0.85 & 0.94 & 0.91& \textbf{0.98} & 0.88 & 0.72 & 0.81  \\
DIA & 0.89 & 0.78 & \textbf{0.93} & \textbf{0.93}& \textbf{0.94} & 0.92 & 0.92 & 0.92& \textbf{0.91} & 0.86 & 0.69 & 0.71  \\
DWA & \textbf{0.92} & 0.9 & \textbf{0.92} & 0.91& \textbf{0.93} & 0.9 & 0.86 & 0.88& \textbf{0.97} & 0.95 & 0.76 & 0.88\\
\hline
\end{tabular}
\caption{Sparsity evaluation on counterfactual explanations.}
\label{exp:sparse}
\end{table*}

\begin{table*}[ht]
\centering
\scriptsize
\begin{tabular}{|c||c|c|c|c||c|c|c|c||c|c|c|c|}

\cline{2-13}
\multicolumn{1}{c||}{} & \multicolumn{4}{c||}{DeepER}
& \multicolumn{4}{c||}{DeepMatcher}
& \multicolumn{4}{c|}{{\ditto}}
\\ \hline
\hline
Dataset & \ourmethod{} & DiCE & SHAP-C & LIME-C & \ourmethod{} & DiCE & SHAP-C & LIME-C & \ourmethod{} & DiCE & SHAP-C & LIME-C \\
\hline
\hline
AB& \textbf{0.54} & 0.45 & 0 & 0& \textbf{0.61} & 0.44 & 0.1 & 0.1& \textbf{0.53} & 0.3 & 0.17 & 0.04\\
AG& 0.41 & \textbf{0.51} & 0.1 & 0.1& 0.54 & \textbf{0.64} & 0.1 & 0.1& \textbf{0.46} & 0.29 & 0.14 & 0.1\\
BA& 0.38 & \textbf{0.49} & 0.12 & 0& 0.31 & \textbf{0.52} & 0 & 0.01& \textbf{0.37} & 0.22 & 0.1 & 0.12 \\
DA& \textbf{0.33} & 0.05 & 0 & 0& \textbf{0.65} & 0.51 & 0 & 0& 0.43 & \textbf{0.44} & 0.03 & 0.05\\
DS&  0.39 & \textbf{0.41} & 0.05 & 0 & \textbf{0.67} & 0.53 & 0 & 0& \textbf{0.31} & 0.29 & 0.01 & 0.05\\
FZ& \textbf{0.35} & 0.31 & 0 & 0& 0.45 & \textbf{0.55} & 0 & 0& 0.34 & \textbf{0.38} & 0.1 & 0.13 \\
IA& \textbf{0.31} & 0.29 & 0 & 0& \textbf{0.8} & 0.29 & 0.24 & 0& 0.12 & 0.04 & 0.13 & \textbf{0.14} \\
WA& \textbf{0.39} & 0.38 & 0 & 0& \textbf{0.56} & 0.5 & 0 & 0& 0.38 & \textbf{0.41} & 0.05 & 0.01 \\
DDA&  \textbf{0.38} & 0.36 & 0.04 & 0& \textbf{0.49} & \textbf{0.49} & 0.1 & 0& \textbf{0.39} & 0.28 & 0.01 & 0.04 \\
DDS & \textbf{0.39} & 0.31 & 0 & 0& \textbf{0.63} & 0.55 & 0.11 & 0& \textbf{0.35} & 0.23 & 0.08 & 0.09 \\
DIA & \textbf{0.41} & 0.35 & 0 & 0& 0.54 & \textbf{0.65} & 0.05 & 0& \textbf{0.46} & 0.19 & 0.12 & 0.19 \\
DWA & \textbf{0.39} & 0.34 & 0 & 0.01& 0.48 & \textbf{0.56} & 0.01 & 0& 0.37 & \textbf{0.45} & 0.15 & 0.03\\
\hline
\end{tabular}
\caption{Diversity evaluation on counterfactual explanations.}
\label{exp:diverse}
\end{table*}

\begin{figure}[!]
    \centering
    \includegraphics[width=0.8\columnwidth]{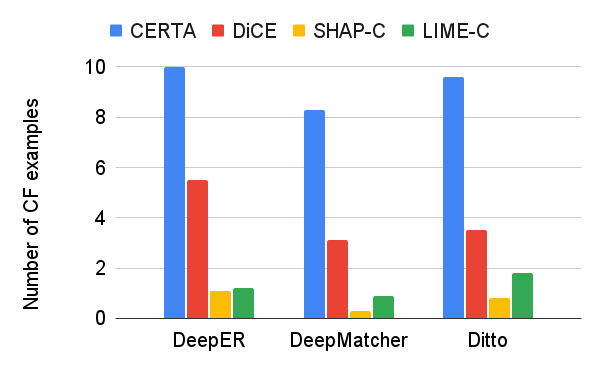}
    \caption{Average number of CF examples generated by CF methods across all considered classifiers and datasets.}
    \label{fig:cfsize1}
\end{figure}

\myparagraph{Counterfactual explanations}
\label{sec:exp-cf}
In Table~\ref{exp:prox} we report the evaluation of \ourmethod{} and baselines for the proximity metric.
For the the DeepER model \ourmethod{} reports better proximity values in $9$ out of $12$ datasets, in the $3$ remaining cases DiCE reports the best proximity value.
In the case of the DeepMatcher model there's a slightly less clear winner, \ourmethod{} and DiCE reach the best proximity on almost the same number of datasets ($6$ for \ourmethod{}, $4$ for DiCE) while they reach a tie on one dataset. On the WA dataset SHAP-C and LIME-C reach the highest proximity, whereas SHAP-C wins on the DS dataset.
Finally, \ourmethod{} reports best proximity on all but one datasets for the \ditto{} classifier, where LIME-C reaches a higher proximity for the IA dataset.
In Table~\ref{exp:sparse} we report the evaluation of \ourmethod{} and baselines for the sparsity metric.
For the DeepER case, \ourmethod{} reports the best sparsity on $6$ datasets out of $12$, a tie is reached between \ourmethod{} and SHAP-C on the BA and DWA datasets. LIME-C reaches the highest sparsity on the AG and DDS datasets.
\ourmethod{} reaches the highest sparsity measure on all datasets, when adopting the DeepMatcher classifier. There are still a couple of ties with DiCE (FZ dataset) and SHAP-C (DDA).  
For the \ditto{} classifier, \ourmethod{} achieves the highest sparsity on $8$ out of $12$ datasets, a tie is reached on the iTunes-Amazon, involving both DiCE and SHAP-C. Another tie involves \ourmethod{} and DiCE for the DS dataset.
In Table~\ref{exp:diverse} we report the evaluation of \ourmethod{} and baselines for the diversity metric.
Across all datasets and models, \ourmethod{} and DiCE reach the best diversity measure, except for the IA case with the \ditto{} classifier.
For DeepER \ourmethod{} gets the highest diversity on $9$ out of $12$ datasets, DiCE instead provides more diverse counterfactual explanations for BA, AG and DS datasets.
On the DeepMatcher classifier \ourmethod{} gets the highest diversity for $6$ datasets, DiCE does the same on $5$ datasets, while they obtain a tie on the remaining dataset (DDA).
Finally, in Figure~\ref{fig:cfsize1} we report the average number of counterfactual explanations generated by \ourmethod{} and baselines for the three considered classifiers.
\ourmethod{} is capable of generating more counterfactual explanations for all the models. Note also that SHAP-C and LIME-C are sometimes not able to generate even a single explanation, as a result the mean number of explanations is below $1$ with SHAP-C for both DeepER and DeepMatcher.

\subsection{Impact of number of triangles}
\label{sec:trimpact}

\ourmethod{} relies on the use of open triangles in order to identify different ways to perturb the records in the original prediction to explain and calculate the probability of sufficiency and necessity associated to the changed attributes.

\begin{figure*}[!]
    \centering
    \subfigure[Probability of sufficiency. \label{fig:tri1}]{
    \includegraphics[width=0.23\textwidth]{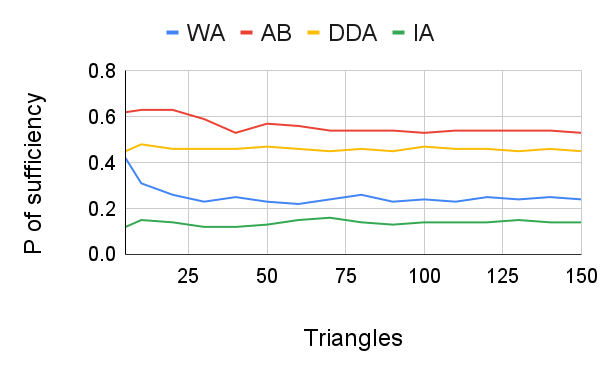}
    }
    \subfigure[Probability of necessity. \label{fig:tri2}]{
    \includegraphics[width=0.23\textwidth]{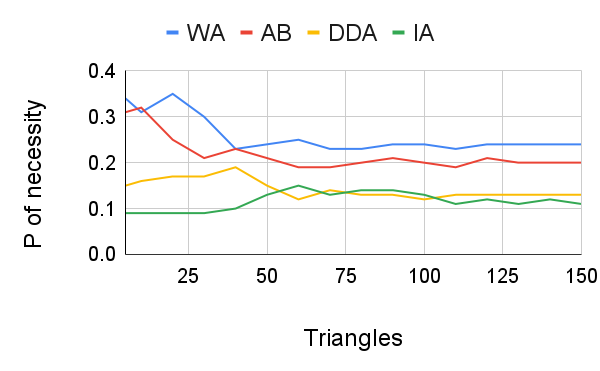}}
    \subfigure[Confidence indication. \label{fig:tri3}]{
    \includegraphics[width=0.23\textwidth]{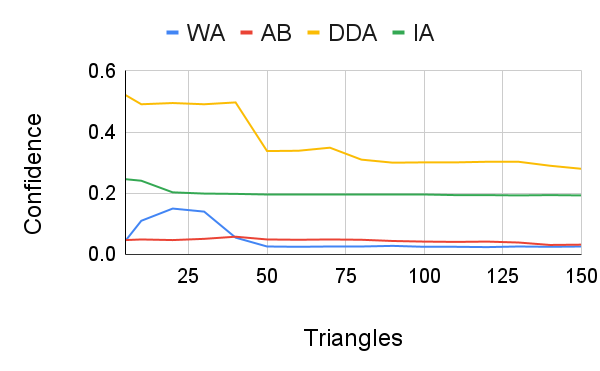}
    }
    \subfigure[Faithfulness. \label{fig:tri4}]{
    \includegraphics[width=0.23\textwidth]{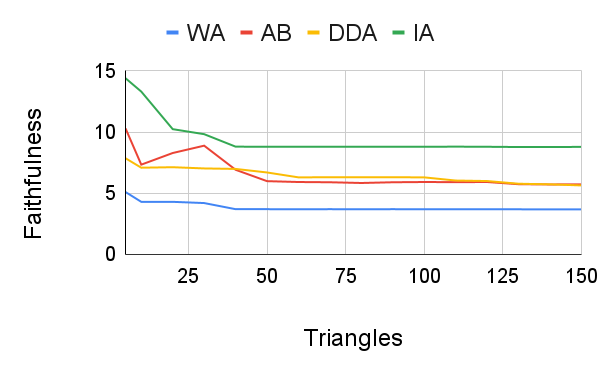}
    }
    \subfigure[Proximity. \label{fig:tri5}]{
    \includegraphics[width=0.23\textwidth]{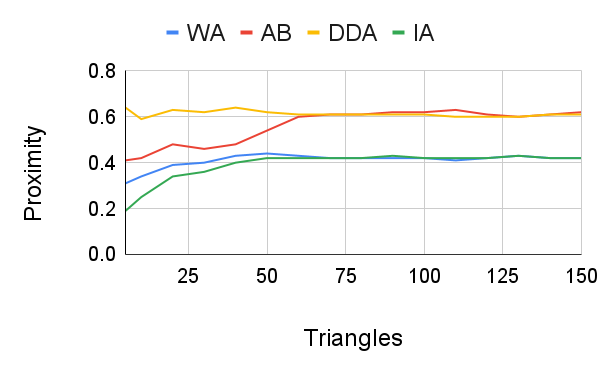}}
    \subfigure[Sparsity. \label{fig:tri6}]{
    \includegraphics[width=0.23\textwidth]{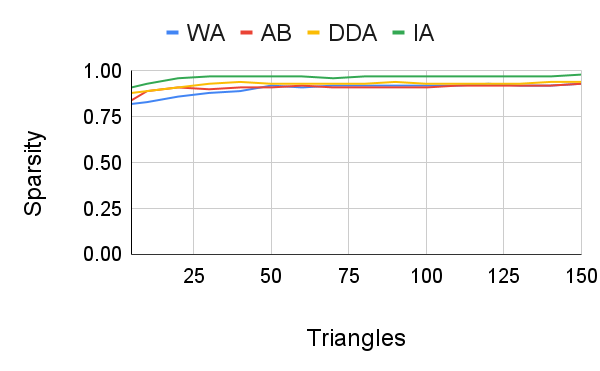}}
    \subfigure[Diversity.  \label{fig:tri7}]{
    \includegraphics[width=0.23\textwidth]{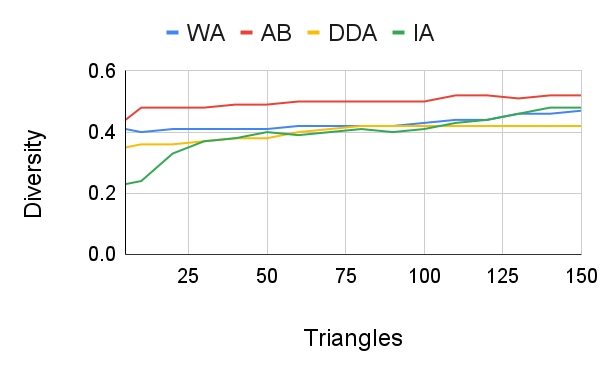}}
    \caption{Probability of Sufficiency (a), Probability of Necessity (b), Confidence Indication (c), Faithfulness (d), Proximity (e), Sparsity (f), Diversity (g) averages, as number of triangles increases.}
\end{figure*}

In this section we study the impact of the number of open triangles adopted in \ourmethod{} along different perspectives. We report how the number of open triangles influences:

\begin{myitemize}
    \item the average probability of sufficiency of a set of attributes in Figure~\ref{fig:tri1};
    \item the average probability of necessity of an attribute in Figure~\ref{fig:tri2};
    \item the quantitative metrics reported in Section~\ref{sec:exp-saliency} for saliency explanations in Figure~\ref{fig:tri3} and Figure~\ref{fig:tri4};
    \item the quantitative metrics reported in Section~\ref{sec:exp-cf} for counterfactual explanations in in Figure~\ref{fig:tri5}, Figure~\ref{fig:tri6} and  Figure~\ref{fig:tri7}.
\end{myitemize}

The evaluations are performed on all three classifiers (DeepER, DeepMatcher and \ditto) on four different datasets (WA, AB, DDA, IA). The results show the average of the reported measure across the three classifiers, for each dataset.

Each of the reported measures in this study tends to converge as the  number of triangles used increases. More specifically, we observe that when \ourmethod{} uses more than $75$-$80$ triangles, it has a generally stable behavior on all the reported metrics. The only metric that increases steadily with the number of triangles is \emph{Diversity} on the DDA and IA datasets.


\subsection{Evaluation of monotonicity assumption}
\label{sec:monotoneeval}

In Section~\ref{sec:lattices} we described how \ourmethod{} builds on the monotone classifier assumption from \cite{tao2018entity} in order to perform as few predictions as possible while tagging the nodes of the lattice structures.
With such an assumption, a flip for a set of attributes $A$ is expected to be propagated in any superset of $A^{\prime} \supset A$. As soon as \ourmethod{} finds a flip $\overline{y}$ for a given $A$, it stops exploring all the upward nodes $A^{\prime} \supset A$, hence the outcomes for any such $A^{\prime}$ are assumed to be $\overline{y}$ without being computed.

Assuming a flip for a set of attributes $A$ induces a flip in any superset $A^{\prime} \supset A$ might overestimate both the probability of sufficiency of $A^{\prime}$ and the probability of necessity of all attributes $a_i \in A$, in case any such predictions for $A^{\prime}$ doesn't result in an actual flip.
This might happen if, following the example in Figure~\ref{fig:lattice-single}, perturbing the value of the attribute $Name_{Abt}$ in the lattice results in a flip, while perturbing the values of the attributes $Name_{Abt}$ and $Price_{Abt}$ doesn't result in a flipped outcome (whereas in Figure~\ref{fig:lattice-single}, where the monotonicity assumption holds, the prediction flips in both cases).

We conduct an experiment to quantify, for a given lattice, how many predictions we save on average, as compared with the number of mistakes we do by assuming monotone classification. To do so we run \ourmethod{} with and without such an optimization and compare the actual outcomes for all the predictions with the case where predictions are propagated based on monotonicity.
We report, for a given lattice:
\begin{myitemize}
    \item the number $l$ of attributes associated to the lattice (\emph{Attributes});
    \item the number of predictions \ourmethod{} needs to make without computing probabilities exactly (\emph{Expected}, equals to $2^{l} - 2$);
    \item the number of predictions performed when \ourmethod{} assumes monotone classification (\emph{Performed});
    \item the number of predictions saved by \ourmethod{} when assuming monotone classification ($\emph{Saved} = \emph{Expected} - \emph{Performed}$);
    \item the ratio between the number of predictions whose \emph{monotone} outcome is different from the actual outcome and the number of saved predictions (\emph{Error rate}).
\end{myitemize}

In Table~\ref{tab:monot} we report the average number of such measures for a given lattice with all classifiers mentioned in Section~\ref{sec:setup}, on four different datasets.

\begin{table}[ht]
\centering
\scriptsize
\begin{tabular}{|c|c|c|c|c|c|}
\hline
\textbf{Dataset} & \textbf{Attributes} & \textbf{Expected} & \textbf{Performed} & \textbf{Saved} & \textbf{Error rate} \\
\hline
\hline
AB      & 3          & 6        & 3.03         & 2.97     & 0.01    \\
\hline
BA      & 3          & 6        & 2.93         & 3.07     & 0.04    \\
\hline
WA      & 4          & 14       & 6.04         & 7.96    & 0.03    \\
\hline
DDS     & 4          & 14       & 4.68         & 9.32    & 0.04    \\
\hline
IA      & 8          & 254      & 45.19        & 208.81   & 0.04  \\
\hline
\end{tabular}
\caption{Average number of expected, performed, saved and wrong predictions on a single lattice.}
\label{tab:monot}
\end{table}

Our comparison reveals the monotone classification assumption allows \ourmethod{} to save $\sim 50\%$ of the predictions for small sets of attributes (AB, BA datasets), with a relatively small error rate, between $1-4\%$. With slightly bigger sets of attributes (WA, DDA datasets) \ourmethod{} saves between $ 57-64\%$ of the predictions, with an error rate between $3-4\%$. The best gain is seen with bigger sets of attributes (IA dataset), where \ourmethod{} saves $\sim 78\%$ of the predictions with an error rate of $\sim 4 \%$.
From our empirical evaluation, the monotone classifier assumption provides an overestimation of the probabilities of at most $4\%$, which seems a reasonable tradeoff especially for bigger sets of attributes, where this allows \ourmethod{} to only perform $\sim 17\%$ of the requested predictions.

\subsection{Impact of data augmentation}
\label{sec:daimpact}

We conduct experiments to quantify the impact of data augmentation mechanism described in Section~\ref{sec:obtaintri} on the effectiveness of \ourmethod{}. We do so by reporting (i) the average number of open triangles \ourmethod{} would generate, without data augmentation (ii) the effect on saliency and counterfactual metrics of forcing the usage of open triangles generated through data augmentation (even when \ourmethod{} could obtain the desired number of open triangles without data augmentation). We run \ourmethod{} with the data augmentation mechanism disabled for BA and FZ datasets using \ditto{} and \emph{DeepMatcher} and report the average number of open triangles generated by \ourmethod{} when targeting $100$ open triangles. In Table~\ref{tab:da-loss} we observe that data augmentation provides \ourmethod{} $10$ to $39\%$ of the requested open triangles.

\begin{table}[!ht]
    \centering
    \begin{tabular}{|c|c|c|}
    \hline
        \textbf{Dataset} & \textbf{DeepMatcher} & \textbf{\ditto{}} \\ \hline \hline
        BA & 90 & 84 \\ \hline
        FZ & 77 & 61 \\ \hline
    \end{tabular}
    \caption{Average number of open triangles generated by \ourmethod{}, when data augmentation is disabled.}
    \label{tab:da-loss}
\end{table}

Additionally we run experiments to report, for each saliency and counterfactual metric, the difference between such a metric value when using exclusively open triangles generated via data augmentation minus the original metric value, when data augmentation is only used in case of open triangle shortage. In Tables~\ref{tab:da-dm} and \ref{tab:da-ditto} we report such a difference, the metrics whose value improve are typed in bold. We observe that \ourmethod{} can benefit from the usage of data augmentation enabled by default, or at most it is not negatively affected.

\begin{table}[!ht]
    \centering
    \begin{tabular}{|c|c|c|c|c|c|}
    \hline
        \textbf{Dataset} & \textbf{Proximity} & \textbf{Sparsity} & \textbf{Diversity} & \textbf{Faithfulness} & \textbf{CI} \\ \hline \hline
        BA & \textbf{0.005} & 0 & \textbf{0.005} & 0 & \textbf{-0.001} \\ \hline
        FZ & \textbf{0.001} & \textbf{0.001} & \textbf{0.006} & 0 & 0 \\ \hline
    \end{tabular}
    \caption{Effect of using open triangles generated via data augmentation on explanation metrics on  DeepMatcher.}
    \label{tab:da-dm}
\end{table}

\begin{table}[!ht]
    \centering
    \begin{tabular}{|c|c|c|c|c|c|}
    \hline
        \textbf{Dataset} & \textbf{Proximity} & \textbf{Sparsity} & \textbf{Diversity} & \textbf{Faithfulness} & \textbf{CI} \\ \hline \hline
        BA & \textbf{0.016} & \textbf{0.002} & \textbf{0.015} & \textbf{-0.046} & \textbf{-0.122} \\ \hline
        FZ & \textbf{0.012} & \textbf{0.002} & \textbf{0.009} & \textbf{-0.004} & \textbf{-0.005} \\ \hline
    \end{tabular}
    \caption{Effect of using open triangles generated via data augmentation on explanation metrics on  \ditto{}.}
    \label{tab:da-ditto}
\end{table}

\subsection{Case Study}

\begin{figure*}[!]
    \centering
    \subfigure[True positive: Label=$1$, Score=$0.67$. \label{fig:cstp}]{
    \includegraphics[width=0.99\textwidth]{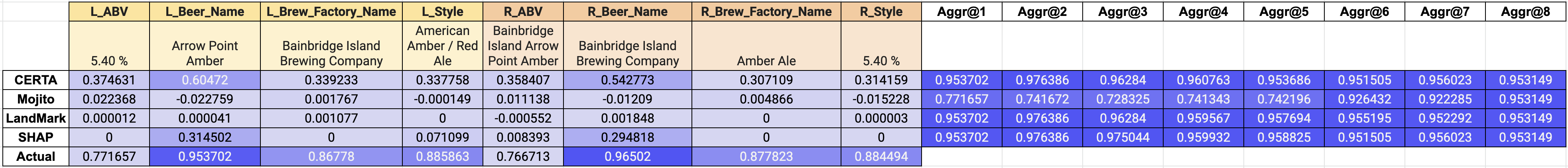}
    }
    \subfigure[True negative: Label=$0$, Score=$0.02$ \label{fig:cstn}]{
    \includegraphics[width=0.99\textwidth]{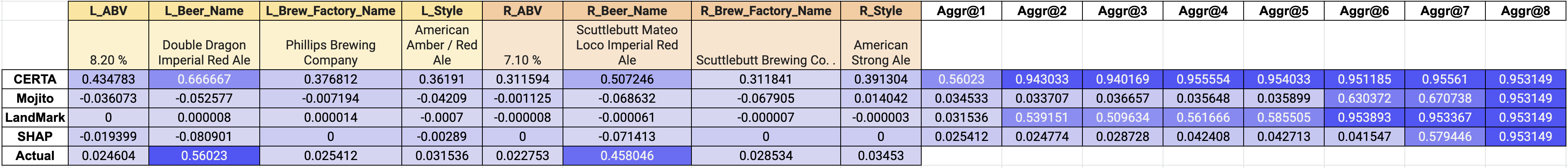}}
    \subfigure[False positive: Label=$0$, Score=$0.51$ \label{fig:csfp}]{
    \includegraphics[width=0.99\textwidth]{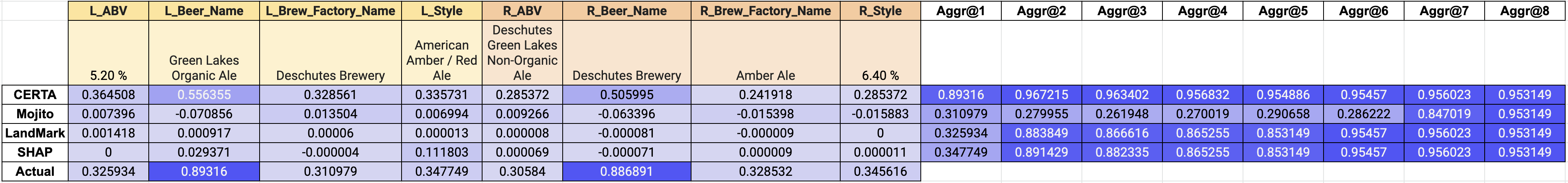}
    }
    \subfigure[False negative: Label=$1$, Score=$0.32$ \label{fig:csfn}]{
    \includegraphics[width=0.99\textwidth]{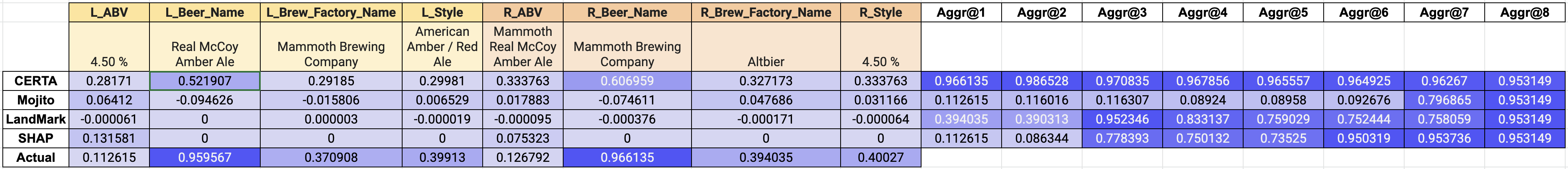}
    }
    \caption{Qualitative analysis on predictions made by \ditto{} on records from BA dataset.}
    \label{fig:cs}
\end{figure*}

In Section~\ref{sec:exp-saliency} we reported the faithfulness for saliency explanations generated by \ourmethod{} and baselines for different datasets and ER systems. In this section we illustrate the superior faithfulness of \ourmethod{} explanations is reflected in single instance by means of a qualitative analysis.
For this sake we let \ditto{} predict a few instances from the test set of the BA dataset, generate the explanations using all the considered saliency methods (see Figure~\ref{fig:cs}). 
As already stated in Section~\ref{sec:exp-method}, when a saliency explanation is good at identifying the attributes that have high influence on a given prediction, modifying attributes with bigger saliency scores should lead to bigger changes in the prediction score, whereas changing attributes with low saliency scores should not alter the score much. 
Given a prediction we report the \emph{effect} of masking the value of each attribute ``in isolation''. Such an effect is measured for each attribute in terms of the difference between the original prediction score and the prediction score of the ER system on the same input with such an attribute masked. Masking an attribute is performed by making the system ignoring its contents. This is referenced in Figure~\ref{fig:cs} as the \emph{Actual} saliency score, which we consider the ``ground truth'' saliency explanation. 
So, ideally, a good explanation should have saliency scores that align with such ground truth saliency scores.
Additionally, we also consider the effect of altering $k$ attributes ``in combination'' by masking the top $k$ salient attributes according to a given saliency explanation and reporting the difference between the prediction scores obtained on the original versus the masked input. This is reported in Figure~\ref{fig:cs} as the \emph{Aggr@}$k$ columns.
Figure~\ref{fig:cs} shows the outcomes of \ourmethod{} and baselines explanation methods for representative predictions computed by \ditto{} on records from BA test set. For easier reading, we prefix attribute names of records $u \in U$ with $L\_$ and $v \in V$ with $R\_$ respectively. 
We observe that for all the cases the top $2$ salient attributes according to the \emph{Actual} saliency (\emph{L\_Beer\_Name}, \emph{R\_Beer\_Name}) coincide with the top $2$ salient attributes according to \ourmethod{}. When looking at the effect of attributes ``in combination'' we observe that the true positive prediction is affected significantly (score difference $> 0.95$) by all explanations, only with Mojito we need to mask $6$ attributes to obtain a score change bigger than $0.9$. In the three remaining examples \ourmethod{} reports the largest effects even for small $k$, when compared with Mojito, LandMark and SHAP. 

\section{Concluding Remarks and Future Work}
\label{sec:conclusion}

In this paper, we introduced the novel \ourmethod{} method for computing saliency and counterfactual explanations for the Entity Resolution (ER) task. Our key insights are the following. (i) Given a pair of records $\langle u,v \rangle$, we identify records $w$ that can form \emph{open triangles}, that is, records from which we can progressively copy values so as to make $\langle u,v \rangle$ less likely to match when initially declared as a match, or more similar when initially declared as a non-match. 
(ii) Given an open triangle $\langle u,v,w \rangle$, we leverage  \emph{lattice data structures} to identify minimal changes to attribute values that can yield a flip in prediction, with few targeted attempts.

Our experimental comparison with baseline solutions demonstrated that \ourmethod{} can find saliency and counterfactual explanations that are more effective on existing deep learning based classifiers, according to established quantitative evaluation metrics.


Future work includes application of \ourmethod{} to other scenarios where the goal is to learn how similar or related two objects are, and thus can display a transitive structure analogous to ER. Examples of such scenarios include schema matching, recommendation systems and handwriting verification. Extension of \ourmethod{}'s principled explanation framework for ER to token-level explanations is another line of future research.


\bibliographystyle{abbrv}
\bibliography{main}

\begin{thebibliography}{10}

\bibitem{arya2019one}
V.~Arya, R.~K. Bellamy, P.-Y. Chen, A.~Dhurandhar, M.~Hind, S.~C. Hoffman,
  S.~Houde, Q.~V. Liao, R.~Luss, A.~Mojsilovi{\'c}, et~al.
\newblock One explanation does not fit all: A toolkit and taxonomy of ai
  explainability techniques.
\newblock {\em arXiv preprint arXiv:1909.03012}, 2019.

\bibitem{DBLP:conf/emnlp/AtanasovaSLA20}
P.~Atanasova, J.~G. Simonsen, C.~Lioma, and I.~Augenstein.
\newblock A diagnostic study of explainability techniques for text
  classification.
\newblock In B.~Webber, T.~Cohn, Y.~He, and Y.~Liu, editors, {\em Proceedings
  of the 2020 Conference on Empirical Methods in Natural Language Processing,
  {EMNLP} 2020, Online, November 16-20, 2020}, pages 3256--3274. Association
  for Computational Linguistics, 2020.

\bibitem{DBLP:conf/edbt/BaraldiBP021}
A.~Baraldi, F.~D. Buono, M.~Paganelli, and F.~Guerra.
\newblock Using landmarks for explaining entity matching models.
\newblock In Y.~Velegrakis, D.~Zeinalipour{-}Yazti, P.~K. Chrysanthis, and
  F.~Guerra, editors, {\em Proceedings of the 24th International Conference on
  Extending Database Technology, {EDBT} 2021, Nicosia, Cyprus, March 23 - 26,
  2021}, pages 451--456. OpenProceedings.org, 2021.

\bibitem{barlaug2021neural}
N.~Barlaug and J.~A. Gulla.
\newblock Neural networks for entity matching: A survey.
\newblock {\em ACM Transactions on Knowledge Discovery from Data (TKDD)},
  15(3):1--37, 2021.

\bibitem{brunner2020entity}
U.~Brunner and K.~Stockinger.
\newblock Entity matching with transformer architectures-a step forward in data
  integration.
\newblock In {\em International Conference on Extending Database Technology,
  Copenhagen, 30 March-2 April 2020}, 2020.

\bibitem{christen2008automatic}
P.~Christen.
\newblock Automatic record linkage using seeded nearest neighbour and support
  vector machine classification.
\newblock In {\em Proceedings of the 14th ACM SIGKDD international conference
  on Knowledge discovery and data mining}, pages 151--159, 2008.

\bibitem{devlin2018bert}
J.~Devlin, M.-W. Chang, K.~Lee, and K.~Toutanova.
\newblock Bert: Pre-training of deep bidirectional transformers for language
  understanding.
\newblock {\em arXiv preprint arXiv:1810.04805}, 2018.

\bibitem{di2019interpreting}
V.~Di~Cicco, D.~Firmani, N.~Koudas, P.~Merialdo, and D.~Srivastava.
\newblock Interpreting deep learning models for entity resolution: an
  experience report using lime.
\newblock In {\em Proceedings of the Second International Workshop on
  Exploiting Artificial Intelligence Techniques for Data Management}, pages
  1--4, 2019.

\bibitem{DBLP:conf/icde/EbaidTAEO19}
A.~Ebaid, S.~Thirumuruganathan, W.~G. Aref, A.~K. Elmagarmid, and M.~Ouzzani.
\newblock {EXPLAINER:} entity resolution explanations.
\newblock In {\em 35th {IEEE} International Conference on Data Engineering,
  {ICDE} 2019, Macao, China, April 8-11, 2019}, pages 2000--2003. {IEEE}, 2019.

\bibitem{ebraheem2018distributed}
M.~Ebraheem, S.~Thirumuruganathan, S.~Joty, M.~Ouzzani, and N.~Tang.
\newblock Distributed representations of tuples for entity resolution.
\newblock {\em PVLDB}, 11(11):1454--1467, 2018.

\bibitem{fellegi1969theory}
I.~P. Fellegi and A.~B. Sunter.
\newblock A theory for record linkage.
\newblock {\em Journal of the American Statistical Association},
  64(328):1183--1210, 1969.

\bibitem{guidotti2018survey}
R.~Guidotti, A.~Monreale, S.~Ruggieri, F.~Turini, F.~Giannotti, and
  D.~Pedreschi.
\newblock A survey of methods for explaining black box models.
\newblock {\em ACM computing surveys (CSUR)}, 51(5):1--42, 2018.

\bibitem{jacovi2020towards}
A.~Jacovi and Y.~Goldberg.
\newblock Towards faithfully interpretable nlp systems: How should we define
  and evaluate faithfulness?
\newblock {\em arXiv preprint arXiv:2004.03685}, 2020.

\bibitem{kommiya2021towards}
R.~Kommiya~Mothilal, D.~Mahajan, C.~Tan, and A.~Sharma.
\newblock Towards unifying feature attribution and counterfactual explanations:
  Different means to the same end.
\newblock In {\em Proceedings of the 2021 AAAI/ACM Conference on AI, Ethics,
  and Society}, pages 652--663, 2021.

\bibitem{lapuschkin2019unmasking}
S.~Lapuschkin, S.~W{\"a}ldchen, A.~Binder, G.~Montavon, W.~Samek, and K.-R.
  M{\"u}ller.
\newblock Unmasking clever hans predictors and assessing what machines really
  learn.
\newblock {\em Nature communications}, 10(1):1--8, 2019.

\bibitem{DBLP:journals/pvldb/0001LSDT20}
Y.~Li, J.~Li, Y.~Suhara, A.~Doan, and W.~Tan.
\newblock Deep entity matching with pre-trained language models.
\newblock {\em Proc. {VLDB} Endow.}, 14(1):50--60, 2020.

\bibitem{DBLP:conf/pkdd/LooverenK21}
A.~V. Looveren and J.~Klaise.
\newblock Interpretable counterfactual explanations guided by prototypes.
\newblock In N.~Oliver, F.~P{\'{e}}rez{-}Cruz, S.~Kramer, J.~Read, and J.~A.
  Lozano, editors, {\em Machine Learning and Knowledge Discovery in Databases.
  Research Track - European Conference, {ECML} {PKDD} 2021, Bilbao, Spain,
  September 13-17, 2021, Proceedings, Part {II}}, volume 12976 of {\em Lecture
  Notes in Computer Science}, pages 650--665. Springer, 2021.

\bibitem{DBLP:conf/nips/LundbergL17}
S.~M. Lundberg and S.~Lee.
\newblock A unified approach to interpreting model predictions.
\newblock In I.~Guyon, U.~von Luxburg, S.~Bengio, H.~M. Wallach, R.~Fergus,
  S.~V.~N. Vishwanathan, and R.~Garnett, editors, {\em Advances in Neural
  Information Processing Systems 30: Annual Conference on Neural Information
  Processing Systems 2017, December 4-9, 2017, Long Beach, CA, {USA}}, pages
  4765--4774, 2017.

\bibitem{martens2014explaining}
D.~Martens and F.~Provost.
\newblock Explaining data-driven document classifications.
\newblock {\em MIS quarterly}, 38(1):73--100, 2014.

\bibitem{DBLP:conf/fat/MothilalST20}
R.~K. Mothilal, A.~Sharma, and C.~Tan.
\newblock Explaining machine learning classifiers through diverse
  counterfactual explanations.
\newblock In M.~Hildebrandt, C.~Castillo, L.~E. Celis, S.~Ruggieri, L.~Taylor,
  and G.~Zanfir{-}Fortuna, editors, {\em FAT* '20: Conference on Fairness,
  Accountability, and Transparency, Barcelona, Spain, January 27-30, 2020},
  pages 607--617. {ACM}, 2020.

\bibitem{mudgal2018deep}
S.~Mudgal, H.~Li, T.~Rekatsinas, A.~Doan, Y.~Park, G.~Krishnan, R.~Deep,
  E.~Arcaute, and V.~Raghavendra.
\newblock Deep learning for entity matching: A design space exploration.
\newblock In {\em Proceedings of the 2018 International Conference on
  Management of Data}, pages 19--34, 2018.

\bibitem{pouyanfar2018survey}
S.~Pouyanfar, S.~Sadiq, Y.~Yan, H.~Tian, Y.~Tao, M.~P. Reyes, M.-L. Shyu, S.-C.
  Chen, and S.~Iyengar.
\newblock A survey on deep learning: Algorithms, techniques, and applications.
\newblock {\em ACM Computing Surveys (CSUR)}, 51(5):1--36, 2018.

\bibitem{primpeli2020profiling}
A.~Primpeli and C.~Bizer.
\newblock Profiling entity matching benchmark tasks.
\newblock In {\em Proceedings of the 29th ACM International Conference on
  Information \& Knowledge Management}, pages 3101--3108, 2020.

\bibitem{DBLP:journals/pvldb/QianPS19}
K.~Qian, L.~Popa, and P.~Sen.
\newblock Systemer: {A} human-in-the-loop system for explainable entity
  resolution.
\newblock {\em Proc. {VLDB} Endow.}, 12(12):1794--1797, 2019.

\bibitem{DBLP:journals/adac/RamonMPE20}
Y.~Ramon, D.~Martens, F.~J. Provost, and T.~Evgeniou.
\newblock A comparison of instance-level counterfactual explanation algorithms
  for behavioral and textual data: Sedc, {LIME-C} and {SHAP-C}.
\newblock {\em Adv. Data Anal. Classif.}, 14(4):801--819, 2020.

\bibitem{ribeiro2016should}
M.~T. Ribeiro, S.~Singh, and C.~Guestrin.
\newblock "why should i trust you?" explaining the predictions of any
  classifier.
\newblock In {\em Proceedings of the 22nd ACM SIGKDD international conference
  on knowledge discovery and data mining}, pages 1135--1144, 2016.

\bibitem{ribeiro2018anchors}
M.~T. Ribeiro, S.~Singh, and C.~Guestrin.
\newblock Anchors: High-precision model-agnostic explanations.
\newblock In {\em Thirty-Second AAAI Conference on Artificial Intelligence},
  2018.

\bibitem{DBLP:journals/corr/abs-1910-01108}
V.~Sanh, L.~Debut, J.~Chaumond, and T.~Wolf.
\newblock Distilbert, a distilled version of {BERT:} smaller, faster, cheaper
  and lighter.
\newblock {\em CoRR}, abs/1910.01108, 2019.

\bibitem{DBLP:journals/access/StepinACP21}
I.~Stepin, J.~M. Alonso, A.~Catal{\'{a}}, and M.~Pereira{-}Fari{\~{n}}a.
\newblock A survey of contrastive and counterfactual explanation generation
  methods for explainable artificial intelligence.
\newblock {\em {IEEE} Access}, 9:11974--12001, 2021.

\bibitem{tao2018entity}
Y.~Tao.
\newblock Entity matching with active monotone classification.
\newblock In {\em Proceedings of the 37th ACM SIGMOD-SIGACT-SIGAI Symposium on
  Principles of Database Systems}, pages 49--62, 2018.

\bibitem{thirumuruganathan2019explaining}
S.~Thirumuruganathan, M.~Ouzzani, and N.~Tang.
\newblock Explaining entity resolution predictions: Where are we and what needs
  to be done?
\newblock In {\em Proceedings of the Workshop on Human-In-the-Loop Data
  Analytics}, pages 1--6, 2019.

\bibitem{DBLP:journals/corr/abs-2010-10596}
S.~Verma, J.~P. Dickerson, and K.~Hines.
\newblock Counterfactual explanations for machine learning: {A} review.
\newblock {\em CoRR}, abs/2010.10596, 2020.

\bibitem{DBLP:journals/corr/abs-1711-00399}
S.~Wachter, B.~D. Mittelstadt, and C.~Russell.
\newblock Counterfactual explanations without opening the black box: Automated
  decisions and the {GDPR}.
\newblock {\em CoRR}, abs/1711.00399, 2017.

\bibitem{wang2016database}
W.~Wang, M.~Zhang, G.~Chen, H.~Jagadish, B.~C. Ooi, and K.-L. Tan.
\newblock Database meets deep learning: Challenges and opportunities.
\newblock {\em ACM SIGMOD Record}, 45(2):17--22, 2016.

\bibitem{wang2018explaining}
X.~Wang, L.~Haas, and A.~Meliou.
\newblock Explaining data integration.
\newblock {\em Data Engineering Bulletin}, 41(2), 2018.

\bibitem{watson2021local}
D.~Watson, L.~Gultchin, A.~Taly, and L.~Floridi.
\newblock Local explanations via necessity and sufficiency: unifying theory and
  practice.
\newblock {\em arXiv preprint arXiv:2103.14651}, 2021.

\end{thebibliography}

\end{document}